\documentclass[lettersize,journal]{IEEEtran}
\usepackage{amsmath,amsfonts}
\usepackage{algorithmic}
\usepackage{algorithm}
\usepackage{array}
\usepackage[caption=false,font=normalsize,labelfont=sf,textfont=sf]{subfig}
\usepackage{textcomp}
\usepackage{stfloats}
\usepackage{url}
\usepackage{verbatim}
\usepackage{graphicx}
\usepackage{cite}
\usepackage{tikz}
\usepackage[all,pdf]{xy}
\usepackage{multirow}
\usepackage{hyperref}
\usepackage{algorithm}
\usepackage{algorithmic}
\usepackage{makecell}

\newtheorem{Lemma}{Lemma}
\newtheorem{Definition}{Definition}
\newtheorem{theorem}{Theorem}
\newtheorem{Example}{Example}

\numberwithin{equation}{section}

\def\be{\begin{equation}}
	\def\ee{\end{equation}}
\def\ba{\begin{array}}
	\def\ea{\end{array}}
\def\RR{{\mathcal R}}
\def\BP{\boldsymbol{\Phi}}
\def\ss{\scriptscriptstyle}
\def\A{\boldsymbol{A}}
\def\B{\boldsymbol{B}}
\def\C{\boldsymbol{C}}
\def\e{\boldsymbol{e}}
\def\m{\boldsymbol{m}}
\def\s{\boldsymbol{s}}
\def\z{\boldsymbol{z}}
\def\a{\boldsymbol{a}}
\def\v{\boldsymbol{v}}
\def\g{\boldsymbol{g}}

\def\Var{{\rm Var}}
\def\c{\boldsymbol{c}}
\def\I{\boldsymbol{I}}
\def\ord{{\rm order}}
\def\ds{\displaystyle}
\def\CMux{{\rm CMux}}

\begin{document}

\title{Large-Plaintext Functional Bootstrapping
	in FHE with Small Bootstrapping Keys}

\author{Dengfa~Liu,
	Hongbo~Li
	\IEEEcompsocitemizethanks{\IEEEcompsocthanksitem
		AMSS, UCAS,
		Chinese Academy of Sciences, Beijing 100080, China.\protect\\
		E-mail: liudengfa19@mails.ucas.ac.cn\protect, hli@mmrc.iss.ac.cn\protect\\}
\thanks{Manuscript received April 19, 2021; revised August 16, 2021.}}

\markboth{Journal of \LaTeX\ Class Files,~Vol.~14, No.~8, August~2021}%
{Shell \MakeLowercase{\textit{et al.}}: A Sample Article Using IEEEtran.cls for IEEE Journals}


\maketitle

\begin{abstract}
Functional bootstrapping is a core technique in Fully Homomorphic Encryption
(FHE). For large plaintext, to evaluate a general function
homomorphically over a ciphertext, 
in the FHEW/TFHE approach, since the function in look-up table form
is encoded in the coefficients of a test polynomial, the degree of the polynomial
must be high enough to hold the entire table. 
This increases the bootstrapping
time complexity and memory cost, 
as the size of bootstrapping keys and keyswitching keys need to be large accordingly.

In this paper, we propose to encode the look-up table of any
function in a polynomial vector, whose coefficients can hold more data. 
The corresponding representation of the additive group
${\mathbb Z}_q$ used in the RGSW-based bootstrapping
is the group of monic monomial permutation matrices, 
which integrates the permutation matrix
representation used by Alperin-Sheriff and Peikert in 2014,
and the monic monomial representation used in the FHEW/TFHE scheme.
We make comprehensive investigation of the new representation, and propose 
a new bootstrapping algorithm based on it.
The new algorithm has the prominent benefit of small bootstrapping key size
and small key-switching key size,
which leads to polynomial factor improvement in key size,
in addition to constant factor improvement in run-time cost. 
\end{abstract}


\section{Introduction}
\IEEEPARstart{F}ully homomorphic encryption (FHE) 
schemes allow to perform arbitrary computation on ciphertexts without resorting to
decryption. After over a decade, FHE schemes have been well developed,
and have been applied to various privacy-preserving tasks, such as private 
database query \cite{tan2020efficient}, private information 
retrieval \cite{yi2012single,angel2018pir}, private decision 
tree evaluation \cite{cong2022sortinghat}, etc.

In FHE schemes,
ciphertexts must be {\it bootstrapped}  
periodically in order to 
run on them arbitrary circuits of arbitrary depth. Originally 
the term referred to refreshing a ciphertext that has a large
error so that after the refreshment, the error decreases to
a small bound \cite{gentry2009fully}. Later on it was 
extended to ``functional bootstrapping", which refers to
evaluating an arbitrary function that is represented by a look-up table. 
Bootstrapping is the most important part of
FHE schemes. 

All bootstrapping schemes are based on Gentry's idea of 
running the decryption circuit homomorphically in a new FHE environment.
\cite{gentry2009fully}.
Take as an example the most popular FHE scheme BFV, whose ciphertexts are
of either LWE or its ring variant RLWE format. The decryption of an LWE ciphertext
consists of two steps: the first step is to compute the {\it phase} of the
ciphertext homomorphically, where the phase is
$\Delta m+e\in {\mathbb Z}_q$, with 
$m\in {\mathbb Z}_t$ being the plaintext, $e$ being the error, and $|e|<\Delta/2$
where $\Delta=\lfloor q/t\rfloor$. 
The second step is to remove $e$ from the 
plaintext in the encrypted phase. The first step is generally easy, while
the second step is difficult.

The trick lies in how to represent the additive group ${\mathbb Z}_q$
in the plaintext space, so that when the group acts on a vector or
polynomial (called {\it test vector} or 
{\it polynomial}) that encodes the look-up table of the function to be
evaluated, the group element corresponding to phase $\Delta m+e$
results in a vector or polynomial with $f(m)$ as its first entry, which 
can then be extracted homomorphically to get a ciphertext encrypting $f(m)$. 

In 2014, Alperin-Sheriff and Peikert \cite{alperin2014faster} proposed a bootstrapping scheme
which we call ``AP14". They use
permutation matrices to represent group ${\mathbb Z}_q$, where
every $u\in [0,q)$ is represented by a $q\times q$ matrix 
\be
\boldsymbol{P}_u:=\left(\ba{cc}
& \I_{u\times u} \\
\I_{(q-u)\times (q-u)}\\
\ea\right), 
\ee
so that the addition of integers agrees with
the multiplication of matrices. By encrypting 
$\boldsymbol{P}_u$ in a GSW-ciphertext \cite{hiromasa2016packing}, 
the homomorphic phase computation
is realized by successive GSW ciphertext multiplications, and the error is small in 
the resulting ciphertext that encrypts the phase of the input ciphertext,
due to the fact that in $\boldsymbol{P}_u$, every row and column respectively  
contain one and only one nonzero entry, and the entry is 1. 

For any function $f$ defined on ${\mathbb Z}_t$, if
extending $f$ to a function $f'$ on ${\mathbb Z}_q$ such that 
$f'(\Delta m+e)=f(m)$ for all $m\in {\mathbb Z}_t$ and $|e|<\Delta/2$, then
for test vector ${\bf test}_f=[f'(0)\ \ f'(1)\ \  \ldots f'(q-1)]$, 
it is easy to verify that
\be\ba{ll}
& {\bf test}_f \times \boldsymbol{P}_{\Delta m+e}\\

=& [f'(\Delta m+e)\ \ f'(\Delta m+e+1)\ \ \ldots\ \  f'(\Delta m+e-1)].
\ea
\label{ap:vector}
\ee
So $f(m)=f'(\Delta m+e)$ is the first entry of the resulting plaintext vector.
(\ref{ap:vector})
is called the {\it test-coefficient look-up property} of the AP14 scheme.

In 2015, Ducas and Micciancio \cite{ducas2015fhew} proposed another bootstrapping 
scheme called ``FHEW". In this scheme, first $q$ is 
switched to $2N$ where $N$ is a power of 2, then every 
$u\in [0,2N)$ is represented by a monic monomial $x^u$, 
so that the addition of integers agrees with
the multiplication of monic monomials. By encrypting every $x^u$
in an RGSW-ciphertext where the polynomial ring is 
${\mathbb Z}[x]/(x^N+1)$, the homomorphic phase computation
is realized by successive RGSW ciphertext multiplications, and the error is small in 
the resulting ciphertext that encrypts the phase of the input ciphertext.

For any function $f$ defined on ${\mathbb Z}_t$, if
$f$ can be extended to a nega-cyclic function $f'$ on ${\mathbb Z}_{2N}$, such that 
$f'(x+N)=-f'(x)$ for all $x\in {\mathbb Z}_{2N}$, and
$f'(\Delta m+e)=f(m)$ for all $m\in {\mathbb Z}_t$ and $|e|<\Delta/2$, 
where $\Delta=\lfloor 2N/t\rfloor$, then
for test polynomial ${\rm test}_f=\sum_{i=0, \ldots, N-1} f'(i)x^{-i}$, 
it is easy to verify that
\be\ba{lll}
{\rm test}_f \times x^{\Delta m+e}
&=& f'(\Delta m+e)+f'(\Delta m+e+1)x^{-1}\\

&& \hfill +\ldots +f'(\Delta m+e-1)x^{-(N-1)}.
\ea
\label{test:FHEW}
\ee
So $f(m)=f'(\Delta m+e)$ is the first entry, a.k.a. the constant term, 
of the resulting plaintext polynomial. (\ref{test:FHEW})
is called the {\it test-coefficient look-up property} of the FHEW scheme.

Since FHEW scheme adopts RGSW ciphertext format in homomorphic phase computation, 
FFT can be used to accelerate the computing. As a result,
bootstrapping a ciphertext encrypting a single-bit plaintext by FHEW scheme
costs less than 1 second.  Later on,
another scheme called TFHE was proposed by Chillotti et
al. in 2016 \cite{chillotti2016faster} to optimize 
homomorphic phase computation in FHEW, which 
improves the run-time cost to less than 0.1 second. In 2021, both schemes were extended to make
functional bootstrapping for arbitrary functions in 2021 \cite{liu2021large,chillotti2021improved}.

The last few years have witnessed fast developments of bootstrapping algorithms.
To name a few, 
in \cite{chillotti2020tfhe}, a scheme was proposed to evaluate several functions
on the same ciphertext simultaneously by one bootstrapping. In 
\cite{liu2021large, yang2021tota}, to bootstrap ciphertexts encrypting 
long plaintext, a strategy of block-wise bootstrapping from the tail up
was proposed; the strategy can be used for functional bootstrapping where
the function is either the identity function $f(x)=x$, 
or the sign function. In \cite{liu2023batch1,liu2023batch2}, FHEW/TFHE was extended to
bootstrap a ciphertext encrypting several plaintexts in SIMD mode.

We notice that for large-plaintext bootstrapping, 
block-wise bootstrapping works only for some special functions. For a
general function, the look-up table representation of the function demands
big polynomial degree in RLWE/RGSW ciphertexts. The bigger the 
polynomial degree, the slower the computation, and the bigger the memory 
cost due to bootstrapping keys and key-switching keys. 

A natural idea is to resort to a polynomial vector instead of
only a polynomial to encode the look-up table 
of a general function $f$. For example, let $N$ be the maximal value of
ring dimension for efficient computing of RLWE/RGSW ciphertext multiplication.
For an ${\rm LWE}_{n,q}$ ciphertext whose plaintext modulus $t$ is big, 
the ciphertext modulus $q$ must be large, say $q>2N$. Let
\be
r=\lceil q/(2N)\rceil, \ \ \ 
q'=2Nr\geq q.
\label{init:r}
\ee
After modulus switch from $q$ to $q'$, and 
extending $f$ to a function defined on ${\mathbb Z}_{q'}$ such that
$f'(\Delta m+e)=f(m)$ for all $m\in {\mathbb Z}_t$ and $|e|<\Delta/2$, 
where $\Delta=\lfloor q'/t\rfloor$, then $f$ can be encoded to the following
$r$-dimensional polynomial vector:
\be
\begin{pmatrix}
	\scriptstyle	f'(0)+f'(1)x+\cdots+f'(N-1)x^{N-1}\\
	\scriptstyle	f'(N)+f'(N+1)x+\cdots+f'(2N-1)x^{N-1}\\
	\vdots\\
	\scriptstyle	f'((r-1)N)+f'((r-1)N+1)x+\cdots+f'(rN-1)x^{N-1}
\end{pmatrix}
\label{1:vector}
\ee

Now that ${\rm test}_f$ in (\ref{test:FHEW}) is replaced by test vector 
(\ref{1:vector}), the representation space of ${\mathbb Z}_{q'}$ must be
replaced by a group of matrices, where in each matrix, every row and column
respectively contains one and only one nonzero entry, and the entry is
a monic monomial. Such matrices are called {\it monic monomial permutation matrices}
in this paper. A subgroup $G$ of the monic monomial permutation matrices is to be
constructed, so that $G$ is isomorphic to ${\mathbb Z}_{q'}$, and the action of
$G$ on (\ref{1:vector}) by matrix-vector multiplication has the 
property that for any $\Delta m+e\in {\mathbb Z}_{q'}$, its
counterpart $g_{\Delta m+e}\in G$ when acting on (\ref{1:vector}), changes the vector into one
whose first entry has its constant term as 
$f'(\Delta m+e)=f(m)$. The last property is the
{\it test-coefficient look-up property}, and is the counterpart of 
(\ref{ap:vector}) and (\ref{test:FHEW}).

The above analysis outlines the main idea of our work. This paper 
proposes to use monic monomial permutation matrices to represent the additive group
${\mathbb Z}_{q'}$, finds among all its cyclic subgroups those that 
has the test-coefficient look-up property, and designs a new scheme
of FHEW/TFHE style, called \texttt{BootMMPM},
for general functional bootstrapping of LWE ciphertexts 
encrypting large plaintext. 

The new scheme \texttt{BootMMPM} has the feature that 
the size of bootstrapping keys and the size of
key-switching keys
are both irrelevant to $r$. 
This is significant for saving memory cost.
Indeed, in recent years there have been several papers dedicated to reducing 
the size of bootstrapping keys. 
Yongwoo Lee et al. \cite{lee2023efficient} proposed a modification of FHEW 
scheme by making phase accumulation with ring automorphisms, which requires much 
fewer bootstrapping keys on the server. In \cite{kim2021general,kim2023lfhe}, bootstrapping keys 
are packed into small number of transfer keys by the client, and then 
reconstructed by the server upon receival. It reduced the key size on the client's side. 
Such method can be applied to TFHE scheme and the new scheme \texttt{BootMMPM} 
to further reduce the key size both on the client's side and on the server's side.

Compared with the TFHE scheme, the new scheme \texttt{BootMMPM} also has slight improvement in
run time. The following table shows that for
$r={\rm poly}(n)$, the time improvement of \texttt{BootMMPM}
is a constant factor, while the
key-size improvement is a polynomial factor.
\begin{table}[H]
	\renewcommand{\arraystretch}{1.3}
	\centering
	\begin{tabular}{|c|c|c|}
		\hline
		Schemes 		& TFHE & \texttt{BootMMPM} \\
		\hline
		\makecell[c]{phase accumulation\\
			time complexity}& 
		$12(nl_BNr)\log (Nr)$ & $12(nl_BNr)\log N$ \\
		\hline
		\makecell[c]{
			bootstrapping\\
			key size}& $8(nl_BN\log Q)r$ &  $8(nl_BN\log Q)$ \\
		\hline
		
		\makecell[c]{
			key-switching\\
			key size}& \makecell[c]{$(n+1)NB_{\rm KS}l_{\rm KS}$\\ $(\log Q)r$} &  
		\makecell[c]{$(n+1)NB_{\rm KS}l_{\rm KS}$\\ $(\log Q)$} \\
		\hline
	\end{tabular}
\end{table}

This paper reports the following work:

\begin{enumerate}
	\item
	Classification of all monic monomial permutation matrices
	under similar transformations. It turns out that 
	any $r\times r$ monic monomial permutation matrix is 
	similar to a block diagonal matrix, where each block
	in the diagonal is of the form
	\be
	\begin{bmatrix}
		&&&&x^{u_0}\\
		x^{u_1}&&&&\\
		&x^{u_2}&&&\\
		&&\ddots&&\\
		&&&x^{u_{k-1}}&
	\end{bmatrix},
	\label{BP1:form:0}
	\ee
	where $k\geq 1$, and if $k=1$, then (\ref{BP1:form:0}) is just $x^{u_0}$.
	
	\item Computation of the order of any monic monomial permutation matrix.
	Only cyclic subgroups of order $2Nr$ in the group ${\rm MMPM}_r$ of
	$r\times r$ monic monomial permutation matrices are useful in bootstrapping.
	
	\item Classification of cyclic subgroups of ${\rm MMPM}_r$ by their 
	number of orbits when acting on the following set of {\it monic monomial indicator vectors}:
	\be
	{\rm MMIV}_r:
	=\{x^i\mathbf{e}_j\,\big|\,i=0, \ldots, 2N-1,\
	j=1, \ldots, r\},
	\ee
	where $\mathbf{e}_j$ is the $r$-dimensional vector whose $j$-entry is $1$
	while all other entries are $0$.
	Only cyclic subgroups of order $2Nr$ and 1 orbit have test-coefficient look-up 
	property.
	
	\item
	construction of test polynomial vector for any cyclic subgroup of order $2Nr$
	and 1 orbit.
	
	\item
	A new algorithm \texttt{BootMMPM} for functional bootstrapping
	based on the cyclic subgroup of order $2Nr$ and 1 orbit that is
	generated by 
	$\begin{bmatrix}
		& x\\
		\I_{(r-1)\times (r-1)}
	\end{bmatrix}$.
	
	\item
	Implementation of \texttt{BootMMPM} on PALISADE library\cite{palisade}, 
	and experiments of bootstrapping with the identity function $f(x)=x$
	on ciphertexts where the plaintext
	size ranges from 5 bits to 15 bits. 
\end{enumerate}

The content is arranged as follows. 
In Section 2, some terminology, notations, 
and basics of FHEW/TFHE functional bootstrapping are introduced.
In Section 3, theoretical investigation of the monic monomial permutation matrix group
is done. 
In Section 4, the new algorithm \texttt{BootMMPM} is proposed and 
analyzed. In Section 5, experimental results on 
\texttt{BootMMPM} are reported.

\section{Notations and Basics of FHEW/TFHE Functional Bootstrapping}
We first present some notations and terminology:

(1) For any integer $n>0$, denote by $[n]$ the set $\{0,1,\cdots,n-1\}$.

(2) Given a polynomial
$a=a_0+a_1x+\cdots+a_{N-1}x^{N-1}$, denote by 
$\overrightarrow{a}=(a_0,a_1,\cdots,a_{N-1})$ its coefficient
vector.

(3) Given a matrix $\boldsymbol{A}$, denote by $\boldsymbol{A}(i,j)$
its $(i,j)$-th entry.


(4) If $g$ is an element of group $G$, denote by 
$\left\langle g\right\rangle$ the cyclic subgroup generated by $g$.

(5) That a random variable $x\in \mathbb Z$ has zero-centered
{\it subgaussian distribution with variance proxy} ${\rm Var}(x)=\sigma\ll Q$, 
refers to the property that there exists a constant $C>0$, such that 
for all $u>0$, ${\rm Prob}(|x|>u)\leq Ce^{-\sigma u^2}$. In particular, if
$x$ is Gaussian with variance $\sigma$, then it is subgaussian with 
variance proxy $\sigma$.

The {\it heuristic bound} of subgaussian random variable $x$ 
is $H\sqrt{\Var(x)}$, where constant $H=O(1)$ is determined by 
the failure probability of the bound.

(6) Let $n,N$ be both powers of 2, where $n<N$. Let
$q\leq q'<Q$ be three positive integers, where $q'$ is even. For any
$M\in \{n,N\}$ and $p\in \{q,q',Q\}$, set 
\be
\RR_{M}:=\mathbb{Z}[x]/(x^M+1),\ \ \
\RR_{M,p}:=\mathbb{Z}_p[x]/(x^M+1).
\ee

(7) Let $t$ be an even positive integer.
A function $f: {\mathbb Z}_{t}\longrightarrow \mathbb{Z}_{t'}$ is 
said to be {\it nega-cyclic}, if for any $k\in [t/2]$,
$f(k+t/2 \ {\rm mod}\ t)=-f(k)\mod t'$.

Let $2N> t$, and 
let $f': {\mathbb Z}_{2N}\longrightarrow \mathbb{Z}_{t'}$ be a
nega-cyclic function such that for all $m\in {\mathbb Z}_{t}$,
all $e\in {\mathbb Z}_{2N}$ satisfying $2|e|< \lfloor 2N/t\rfloor$,
\be
f'(m\lfloor 2N/t\rfloor+e)=f(m).
\ee
Then $f'$ is called a {\it nega-cyclic extension} of $f$ from
${\mathbb Z}_{t}$ to ${\mathbb Z}_{2N}$.

(8) An ${\rm LWE}_{n,q}$ {\it ciphertext of BFV format},
with modulus $q$, dimension $n$, and secret $\s\in {\mathbb Z}_q^n$,
is of the form $(\a, b)$, where $\a\in {\mathbb Z}_q^n$, 
$b\in {\mathbb Z}_q$, and the phase
\be
{\rm phase}(\a,b):=b-\a\cdot \s=m\lfloor q/t\rfloor+e \ {\rm mod} \ q,
\label{def:phase}
\ee
such that $m\in {\mathbb Z}_t$ is the plaintext, and
$e\in {\mathbb Z}_q$ is the error.
The ciphertext is decryptable if and only if
$2|e| <\lfloor q/t\rfloor$. 

Given a message $m\in {\mathbb Z}_t$, the
{\it trivial encryption} of $m$ in ${\rm LWE}_{n,q}$ is the ciphertext 
$(0^n, m\lfloor q/t\rfloor){\rm mod} \ q$. It is independent of secret $\s$.

(9) Given an ${\rm LWE}_{n,q}$ ciphertext $(\a,b)$, the {\it modulus switch}
from $q$ to $q'$ outputs an ${\rm LWE}_{n,q'}$ ciphertext $(\a',b')$ that
encrypts the same plaintext as the input ciphertext:
\be
(\a',b')=(\lfloor \a q'/q\rceil, \lfloor b q'/q\rceil).
\ee

(10) Let $B>1$ be an integer. The $B$-{\it digit decomposition} (or {\it gadget decomposition})
of an integer $x\in [0,q)$ generates an 
integer sequence of length 
$l_B:=\lceil \log_B q\rceil$: $x_0, x_1, \ldots, x_{l_B}$, where
each $x_i\in [B]$, and $x=\sum_{i\in [l_B]} x_i B^i$.

The $B$-digit decomposition operation on ${\mathbb Z}_q$ is denoted by 
$\g_B^{-1}$. Integer $B$ is called the {\it digit decomposition base}.

(11) Given an ${\rm LWE}_{n,q}$ ciphertext $(\a,b)$ with secret
$\s\in {\mathbb Z}_q^n$, the {\it key switch}
from $\s$ to $\z\in {\mathbb Z}_q^N$ outputs an ${\rm LWE}_{N,q}$ 
ciphertext $(\a_z,b_z)$ encrypting the same plaintext as the input
ciphertext but under secret $\z$. 

Key switch requires not only
all the components of $\s=(s_1, \ldots, s_n)$ to be encrypted
with $\z$ beforehand, but also the multiples of the components
of the form $s_i B_{KS}^j k$, where (i) $i\in \{1,\ldots,n\}$, 
(ii) $B_{\rm KS}$ is the 
digit decomposition base, (iii) $j\in [l_{\rm KS}]$ where
$l_{\rm KS}=\lceil \log_{B_{\rm KS}} q \rceil$,
(iv) $k\in [B_{\rm KS}]$. Let $\texttt{ct}(i,j,k)$
be an ${\rm LWE}_{N,q}$ ciphertext encrypting 
$s_i B_{KS}^j k\ {\rm mod }\ q$ with secret $\z$. 
These ciphertexts are called the {\it key-switching keys}; they cannot be
decrypted correctly with $\z$.

Let $\a=(a_1, \ldots, a_n)$, and for each $a_i$ taken as an integer in
$[0,q)$, let 
$\g^{-1}_{B_{KS}}(a_i)=(a_{i,j})_{j\in [l_{\rm KS}]}$ be its
$B_{KS}$-digit decomposition. Then
\begin{equation}
	(\a_z,b_z)=(0^N,b)-\sum\limits_{i\in \{1,\ldots,n\}, j\in [l_{\rm KS}]} 
	\hskip -.5cm
	\texttt{ct}(i,j,a_{i,j}) \ \ \, {\rm mod }\  q.
	\label{equ-key-swi}
\end{equation}

(12) An ${\rm RLWE}_{n,q}$ {\it ciphertext of BFV format}, 
with modulus $q$, ring dimension $n$ that is a power of 2, and
secret key $s\in {\cal R}_{n,q}$, is of the form $(a, b)$, where 
$a, b\in {\cal R}_{n,q}$, and the phase
\be
{\rm phase}(a,b):=b-as=m\lfloor q/t\rfloor+e\ {\rm mod}\ q,
\label{def:phase2}
\ee
such that $m\in {\cal R}_{n,t}$ is the plaintext, and
$e\in {\cal R}_{n,q}$ is the error.

The ciphertext is decryptable if and only if
$
2\|e\|_\infty<\lfloor q/t\rfloor. 
$
Given a message $m\in {\cal R}_{n,t}$, the
{\it trivial encryption} of $m$ in ${\rm RLWE}_{n,q}$ is the ciphertext 
$(0, m\lfloor q/t\rfloor)$. It is independent of secret $s$.

(13) Given an ${\rm RLWE}_{n,q}$ ciphertext
$(a,b)$ (where $a=\sum_{i\in [n]}a_ix^i$, $b=\sum_{j\in [n]}b_j x^j$)
encrypting a message $m=\sum_{k\in [n]} m_kx^k$ with
secret key $s=\sum_{l\in [n]}s_lx^l$, 
the {\it constant-term extraction}
outputs an ${\rm LWE}_{n,q}$ ciphertext 
$(a'',b'')$ encrypting the constant term $m_0\in {\mathbb Z}_t$ with
secret key $\overrightarrow{s}=(s_0,s_1,\ldots,s_{n-1},1)\in {\mathbb Z}_q^n$:
\begin{equation}
	(a'',b'')=(\overrightarrow{a(x^{-1})},b_0)
	=(a_0,-a_{n-1},\ldots,-a_2,-a_1,b_0).
	\label{equ-sam-extr}
\end{equation}

(14) An ${\rm RGSW}_{n,q}$ {\it ciphertext}
with modulus $q$, ring dimension $n$ that is a power of 2, 
secret key $s\in {\cal R}_{n,q}$, and digit decomposition base $B$, 
is an $2l_B\times 2$ matrix $C$ with entries in ${\cal R}_{n,q}$, 
where $l_B=\lceil\log_B q\rceil$, such that the phase
\be
{\rm phase}(C):=C\left(\ba{c}
1\\
-s
\ea\right)=
m\left(\ba{c}
\g_B\\
-s\g_B
\ea\right)
+\e\ {\rm mod}\ q,
\label{def:phase3}
\ee
with $\g_B=[B^0\ B^1\ \cdots\ B^{l_B-1}]^T\in {\mathbb Z}^{2l_B}$,
$m\in {\cal R}_{n,t}$ being the plaintext, and
$\e\in {\cal R}_{n,q}^{2l_B}$ being the error.

The ciphertext is decryptable if and only if
$
2\|\e\|_\infty<\lfloor q/t\rfloor. 
$
Given a message $m\in {\cal R}_{n,t}$, the
{\it trivial encryption} of $m$ in ${\rm RGSW}_{n,q}$ is the ciphertext 
$m\left(\ba{cc}
\g_B\\
& \g_B
\ea\right)$. It is independent of secret $s$.

The multiplication of an ${\rm RLWE}_{n,q}$ ciphertext $(a,b)$
with an ${\rm RGSW}_{n,q}$ ciphertext $C$ under the same secret, 
is
\be
\left(\g_B^{-1}(a,b) \right) C \ {\rm mod }\ q\ \ \ 
\in {\cal R}_{n,q}^{2},
\label{def:gsw}
\ee
where the $B$-digit decomposition operator $\g_B^{-1}$ acts on 
$a,b$ separately, so that $\g_B^{-1}(a,b)\in {\cal R}_{n,B}^{2l_B}$.
(\ref{def:gsw}) is an ${\rm RLWE}_{n,q}$ ciphertext encrypting the
product of the plaintexts in the two ciphertexts respectively.
Similarly, the multiplication between two ${\rm RGSW}_{n,q}$ 
ciphertexts $C_1, C_2$, is $\left(\g_B^{-1}(C_1)\right)C_2$.

(15) The {\it CMux} (controlled multiple executions) operator controlled
by ternary variable $d\in \{1,0,-1\}$ and three operations
$E_1, E_0, E_{-1}$, outputs operation $E_d$. Set
\be
d_{+}:=\max(d,0), \ \ \ \,
d_{-}:=\max(-d,0).
\label{d:pm}
\ee
Then the CMux operator can be denoted by $\CMux(d+,d-;E_1, E_0, E_{-1})$,
whose expression is
\be
E_0+(E_1-E_0)d_+
+(E_{-1}-E_0)d_-
= \left\{\ba{ll}
E_1, & \hbox{if } d=1, \\
E_0, & \hbox{if } d=0, \\
E_{-1}, & \hbox{if } d=-1.
\ea
\right.
\label{equ-exten-cmux}
\ee

\begin{Definition}
	Given an ${\rm LWE}_{n,q}$ ciphertext encrypting message $m\in {\mathbb Z}_t$,
	and a nega-cyclic function $f: {\mathbb Z}_t\longrightarrow {\mathbb Z}_{t'}$, 
	the functional bootstrapping is a procedure of generating a decryptable
	${\rm LWE}_{n,q}$ ciphertext encrypting message $f(m)\in {\mathbb Z}_{t'}$ with 
	the same secret key as the input.
\end{Definition}

Gentry's bootstrapping idea is to execute the decryption circuit 
homomorphically. As the decryption needs to use the secret key $\s$, it must be
provided beforehand and in a ciphertext form. 
The first step of decryption is to compute the
phase of the input LWE ciphertext $\texttt{ct}_0=(\a,b)$, 
which is essentially the inner product between vectors $\a, \s$. Now that
$\s$ is encrypted, the inner product is between a plaintext vector and
a ciphertext vector. 
This can be done similar to the key switch procedure. 

For example,
for $\a=(a_1, \ldots, a_n)$, let 
$\g^{-1}_{B}(a_i)=(a_{i,j})_{j\in [l_B]}$ be the
$B$-digit decomposition of $a_i$, where $l_B=\lceil \log_B q\rceil$;
if $z\in {\cal R}_{N,Q}$ is the new secret key, then
as in (\ref{equ-key-swi}), for $\s=(s_i)_{i=1,\ldots,n}$, 
let each $a_{i,j} B^j s_i\, {\rm mod }\ Q$ be encrypted
to an ${\rm RGSW}_{N,Q}$ ciphertext $\texttt{ct}(i,j,a_{i,j})$ 
with secret key $\z$, then 
\be
\texttt{ct}_2=b\left(\ba{cc}
\g_B\\
& \g_B
\ea\right)-\sum\limits_{i\in \{1,\ldots,n\}, j\in [l_{\rm KS}]} 
\hskip -.5cm
\texttt{ct}(i,j,a_{i,j}) \ \ \, {\rm mod }\  Q
\label{old:boot}
\ee
is an ${\rm RGSW}_{N,Q}$ ciphertext encrypting
${\rm phase}(\a,b) \ {\rm mod}\ q$. The procedure of
computing the additions in (\ref{old:boot}) homomorphically is called
{\it phase accumulation}. The ciphertexts $\texttt{ct}(i,j,a_{i,j})$ are called
{\it FHEW bootstrapping keys}.

The second step of decryption is to remove the error in the phase
by rounding, and execute the ``modulo-$q$" operation.  In bootstrapping,
both need to be done homomorphically. 
To get rid of the error part of ${\rm phase}(\a,b)\, {\rm mod }\ q$, 
the FHEW scheme uses the monomial representation of ${\mathbb Z}_q$: 
for a power-of-2 integer $N>q$, first
the input ciphertext is changed into an ${\rm LWE}_{n,2N}$ ciphertext 
$\texttt{ct}_1$
by modulus switch, then each $a_{i,j} B^j s_i\, {\rm mod }\ 2N$
is represented as a monic monomial $x^{a_{i,j} B^j s_i}$ before being encrypted
in an ${\rm RGSW}_{N,Q}$ ciphertext. 

In the plaintext,
the additions in phase accumulation
(\ref{old:boot}) are done in the exponent space
of a monomial; in the ciphertext,
the additions are done by RGSW ciphertext multiplications.
Since RGSW ciphertexts have the unique
property that the error in the product of two RGSW ciphertexts is
additive in the error of the second ciphertext, ciphertext $\texttt{ct}_2$
has small error growth.
After phase accumulation, the plaintext becomes 
\be
x^{{\rm phase}(\texttt{ct}_1)}=x^{m\lfloor 2N/t\rfloor+e'}
=x^{m\lfloor 2N/t\rfloor} x^{e'}, 
\label{FHEW:phase}
\ee
where the error $e'$ satisfies $2|e'|<\lfloor 2N/t\rfloor$. 

For a nega-cyclic function $f: {\mathbb Z}_t\longrightarrow 
{\mathbb Z}_{t'}$, let $f': {\mathbb Z}_{2N}\longrightarrow 
{\mathbb Z}_{t'}$ be a nega-cyclic extension of $f$, so that
for all
$u=i\lfloor 2N/t\rfloor+e\ {\rm mod}\ 2N$, where $i\in {\mathbb Z}_t$, and 
$e\in {\mathbb Z}_{2N}$ satisfies $2|e|<\lfloor 2N/t\rfloor$, 
\be
f'(u)=
\sum_{\stackrel{\ss i\in [t], e\in {\mathbb Z}_{2N},}
	{\ss 2|e|<\lfloor 2N/t\rfloor}} f(i)
x^{-i\lfloor 2N/t\rfloor-e}\in {\cal R}_{N,t'}.
\label{FHEW:tpol}
\ee
The right side of (\ref{FHEW:tpol}) is a polynomial in ${\cal R}_{N,t'}$, called
the {\it test polynomial} of $f$. 

In FHEW scheme, the trivial ${\rm RLWE}_{N,Q}$-encryption of test polynomial 
(\ref{FHEW:tpol}) is multiplied with ${\rm RLWE}_{N,Q}$ ciphertext 
$\texttt{ct}_2$, the result is an ${\rm RLWE}_{N,Q}$ ciphertext
$\texttt{ct}_3$ encrypting plaintext 
$f'(u) x^{m\lfloor 2N/t\rfloor} x^{e'}\in {\cal R}_{N,t'}$.

The constant term of the plaintext in $\texttt{ct}_3$
is exactly $f(m)$. Then constant-term extraction
is used to obtain from $\texttt{ct}_3$
an ${\rm LWE}_{N,Q}$ ciphertext $\texttt{ct}_4$
encrypting $f(m)$ with secret key
$\overrightarrow{z}$, according to (\ref{equ-sam-extr}).

After this, the key switch procedure is used to change the secret key
to $\s\in {\mathbb Z}_q^n$, resulting in an ${\rm LWE}_{n,Q}$ 
ciphertext $\texttt{ct}_5$. Finally, the modulus switch from
$Q$ to $q$ changes $\texttt{ct}_5$ to an ${\rm LWE}_{n,q}$ 
ciphertext $\texttt{ct}_6$, which is the output. This is the whole
procedure of the FHEW scheme.

The TFHE scheme improves upon the FHEW scheme by merging the 
generation of $\texttt{ct}_2$ and $\texttt{ct}_3$ 
it starts from 
the trivial ${\rm RLWE}_{N,Q}$-encryption of test polynomial 
(\ref{FHEW:tpol}), and consecutively multiplies it from the right side with 
an ${\rm RGSW}_{N,Q}$ ciphertext each encrypting a term 
of (\ref{old:boot}). Then $\texttt{ct}_3$ is generated without generating
$\texttt{ct}_2$. This trick replaces the product of two matrices to
the product between a matrix and a vector.

A typical setting is where $\s$ is ternary, namely, $\s=(s_i)_{i=1,\ldots,n}$
where each $s_i\in \{1,0,-1\}$. In this setting, 
the plaintext $x^{-a_is_i}$ that corresponds to the term $-a_is_i$ in 
${\rm phase}(\a,b)=b-\sum_{i=1, \ldots, n}a_is_i$,
has three possibilities: $x^{-a_i}, 1, x^{a_i}$, 
if $s_i=1,0,-1$, respectively. In terms of the CMux operator and its expression 
(\ref{equ-exten-cmux}), for $s_{i+}=\max(s_i,0)$ and 
$s_{i-}=\max(-s_i,0)$, 
\be\ba{lll}
x^{-a_is_i} &=& 
\CMux(s_{i+},s_{i-}; x^{-a_i}, 1, x^{a_i})\\

&=& 1+(x^{-a_i}-1)s_{i+}+(x^{a_i}-1)s_{i-}.
\ea
\label{CMux:s}
\ee

In (\ref{CMux:s}), if both
$s_{i+},s_{i-}$ are encrypted as ${\rm RGSW}_{N,Q}$ ciphertexts, and 1 is trivially encrypted,
then the right side becomes the sum of three ${\rm RGSW}_{N,Q}$ ciphertexts; it
replaces the product
$\prod_{j\in [l_{\rm KS}]} \texttt{ct}(i,j,a_{i,j})$
in FHEW phase accumulation (\ref{old:boot}). 
The ${\rm RGSW}_{N,Q}$ ciphertexts encrypting the
$s_{i+},s_{i-}$ are called {\it TFHE bootstrapping keys} when the secret is
ternary.

In summary, the procedure of TFHE functional
bootstrapping for LWE ciphertext with ternary secret is as follows:

Input: 
\begin{enumerate}
	\item ${\rm LWE}_{n,q}$ ciphertext $\texttt{ct}_0$ to be bootstrapped,
	whose plaintext modulus is $t$, whose plaintext is $m\in {\mathbb Z}_t$ and
	whose secret is $\s=(s_i)_{i=1, \ldots,n}$;
	
	\item TFHE bootstrapping keys encrypting $s_{i+},s_{i-}$ for
	$i\in \{1, \ldots, n\}$, which are ${\rm RGSW}_{N,Q}$ ciphertexts whose
	secret is $z=\sum_{i\in [N]} z_ix^i\in {\cal R}_{N,Q}$; 
	
	\item test polynomial of the form (\ref{FHEW:tpol}) in ${\cal R}_{N,t'}$ ,
	which encodes the nega-cyclic function $f: {\mathbb Z}_t\longrightarrow 
	{\mathbb Z}_{t'}$;
	
	\item key-switching keys, which are ${\rm RLWE}_{n,Q}$ ciphertexts encrypting 
	$z_i B_{\rm KS}^j k\ {\rm mod }\ Q$ with secret $\s$ for all $i\in [N]$, 
	$j\in [\lceil \log_{B_{\rm KS}} Q \rceil]$, $k\in [B_{\rm KS}]$, where
	$B_{\rm KS}$ is the digit decomposition base for key switch.
\end{enumerate}

Output: ${\rm LWE}_{n,q}$ ciphertext $\texttt{ct}_5$ encrypting
$f(m)\in {\mathbb Z}_{t'}$.

Step 1. Modulus switch from $q$ to $q'=2N$. The result is an
${\rm LWE}_{n,q'}$ ciphertext $\texttt{ct}_1:=(a_1, \ldots, a_n, b)$
with secret $\s$.

Step 2. Phase accumulation (or {\it blind rotation}). It starts from the product
$\texttt{ct}_2$ of the trivial ${\rm RLWE}_{N,Q}$ encryption of the
test polynomial and the trivial ${\rm RGSW}_{N,Q}$ encryption of
$b\in {\mathbb Z}_q$, for every
$i=1, \ldots, n$, updates $\texttt{ct}_2$ by multiplying it from the right side with an
${\rm RGSW}_{N,Q}$ ciphertext encrypting 
$\CMux(s_{i+},s_{i-}; x^{-a_i}, 1, x^{a_i})$. The result, still denoted by $\texttt{ct}_2$,
is an ${\rm RLWE}_{N,Q}$ ciphertext with secret $z$.

Step 3. Constant-term extraction (also called {\it sample extract})
of $\texttt{ct}_2$. The result is
an ${\rm LWE}_{N,Q}$ ciphertext $\texttt{ct}_3$ encrypting $f(m)$
with secret $\overrightarrow{z}$.

Step 4. Key switch from $\overrightarrow{z}$ to $\s$. The result is
an ${\rm LWE}_{n,Q}$ ciphertext $\texttt{ct}_4$.

Step 5. Modulus switch from $Q$ to $q$. The result is an
${\rm LWE}_{n,q}$ ciphertext $\texttt{ct}_5$.

\section{Monomial Matrix Representation of Additive Cyclic Group}
The {\it order} of an element $g$ in a group, denoted by $\ord(g)$, is the smallest
positive integer $m$ such that $g^m=1$, where 1 is the identity element.

When a group $G$ acts on a set $S$, the action is said to be {\it
	transitive}, if for all $x,y\in S$, there exists a $g\in G$ such that
$gx=y$. The action is said to be {\it regular} or {\it faithful}, if for any $g\in G$, 
$gx=x$ for all $x\in S$ if and only if $g=1$.

An $\mathbb F$-{\it representation} of a group $G$ is a homomorphism from $G$ to some
$GL_{\mathbb F}({\mathcal V})$, where
${\mathcal V}$ is a finite-dimensional $\mathbb F$-vector space.
Two representations $\BP_i: G\longrightarrow GL_{\mathbb F}({\mathcal V}_i)$
for $i=1,2$, are said to be {\it equivalent}, if there is an $\mathbb F$-linear
isomorphism $\boldsymbol{T}: {\mathcal V}_1\rightarrow {\mathcal V}_2$,
such that for all $g\in G$,
$\BP_2(g)= \boldsymbol{T}\BP_1(g)\boldsymbol{T}^{-1}$.

Now fix the field $\mathbb F$ as the following one, where $N$ is a power of 2:
\be
{\mathbb F}=\mathbb{Q}(x)/(x^N+1).
\ee
For any $r>0$,
an element of $GL({\mathbb F}^r)$ is called a {\it monic monomial permutation matrix}
if in the matrix, every row and every column respectively contains one and only
one nonzero entry, and the entry is a monic monomial.
All $r\times r$ monic monomial permutation matrices form a finite subgroup ${\rm MMPM}_r$
of $GL({\mathbb F}^r)$.
It is easy to see that
any $r\times r$ monic monomial permutation matrix $\A$ is of the form
\be
\A=(x^{u_i}\delta_{i,\phi(j)})_{i,j\in [r]},
\label{BP:1}
\ee
where $u_i\in [2N]$,
$\delta$ is the Kronecker symbol, and $\phi\in S_r$ is
a permutation acting on $[r]$.

Consider the additive cyclic group ${\mathbb Z}_{q'}$, where $q'=2Nr$.
The identity element is 0, and the generator is 1.
Let
$\BP:\mathbb{Z}_{q'}\longrightarrow {\rm MMPM}_r$ be a representation
of group $\mathbb{Z}_{q'}$. Then $\BP(\mathbb{Z}_{q'})$ is
a subgroup generated by matrix $\BP(1)$. 

\begin{Lemma}\label{structure:0}
	Matrix $\BP(1)$ is similar to ${\rm diag}(\A_1, \B)$, where if setting $k=1$, then
	$\A_k\in {\rm MMPM}_{r_k}$ with
	$1\leq r_k\leq r$, $\B\in {\rm MMPM}_{r-r_k}$,
	such that
	\be
	\A_k = \begin{bmatrix}
		&&&&x^{u_{k,0}}\\
		x^{u_{k,1}}&&&&\\
		&x^{u_{k,2}}&&&\\
		&&\ddots&&\\
		&&&x^{u_{k,r_k-1}}&
	\end{bmatrix},
	\label{form:Ak}
	\ee
	with each $u_{k,i}\in [2N]$.
\end{Lemma}

\begin{IEEEproof}
	Let $r_1$ be the smallest among all the positive integers $m$ satisfying $\phi^m(0)=0$.
	Then $1\leq r_1\leq r$, and $\{\phi^i(0),\ i\in [r_1]\}$ is
	the orbit of 0 under the action of $\phi$, which contains $r_1$ different elements.
	Choose an arbitrary bijection $\psi:[r]\backslash
	[r_1]\longrightarrow[r]\backslash \{\phi^i(0),\ i\in [r_1]\}$.
	Construct the following element $\psi'\in S_r$: for all $t\in [r]$,
	\be
	\psi'(t)=\left\{
	\ba{ll}
	\phi^t(0), &\hbox{ if } t\in [r_1], \\
	\psi(t), &\hbox{ else.}
	\ea\right.
	\label{def:psi}
	\ee 	
	
	Set matrix $\boldsymbol{T}=\big(\delta_{i,\psi'(j)}\big)_{i,j\in [r]}$.
	Then $\boldsymbol{T}^{-1}=\big(\delta_{\psi'(i),j})_{i,j\in [r]}$.
	For all $i,j\in [r]$, the $(i,j)$-entry of matrix $\boldsymbol{T}^{-1}\BP(1)\boldsymbol{T}$ is
	\be\ba{lll}
	\ds\sum\limits_{k,l\in[r]}
	\delta_{\psi'(i),k} (x^{u_k}\delta_{k,\phi(l)})
	\delta_{l,\psi'(j)}
	&=& x^{u_{\psi'(i)}}\delta_{\psi'(i),\phi\psi'(j)}\\
	
	&=& x^{u_{\psi'(i)}}\delta_{i,{\psi'}^{-1}\phi\psi'(j)}.
	\ea
	\ee
	Set $\phi'={\psi'}^{-1}\phi\psi'$, set $u'_i=u_{\psi'(i)}$, and set
	$u_{1,i}=u'_i$ if $i\in [r_1]$.
	
	For any $j\in [r]$, if
	$j\in [r_1]$, then $\phi'(j)=\phi^j(0)$ by (\ref{def:psi}), so
	$\delta_{i,\phi'(j)}=1$ if and only if
	$i=\phi'(j)={\psi'}^{-1}\phi^{j+1}(0)=j+1\ {\rm mod}\ r_1\in [r_1]$. If
	$j\in [r]\backslash [r_1]$, then $\phi\psi'(j)=\phi\psi(j)\notin
	\{\phi^k(0),\ k\in [r_1]\}$, so
	$\phi'(j)={\psi'}^{-1}\phi\psi'(j)\notin [r_1]$. In particular, $[r]\backslash [r_1]$ is
	invariant under $\phi'$.
	
	So matrix $\boldsymbol{T}^{-1}\BP(1)\boldsymbol{T}$ is block-diagonal,
	the first block in the diagonal is $\A_1$, and the second block is 
	$\B=\left(x^{u'_{i+r_1}}\delta_{i+r_1,\phi'(j+r_1)}\right)_{i,j\in [r-r_1]}$.
\end{IEEEproof}

\begin{Lemma}\label{structure:1}
	Matrix $\BP(1)$
	is similar to ${\rm diag}(\A_1, \A_2, \ldots, \A_h)$, where
	$h\geq 1$, and for every $1\leq k\leq h$,
	$\A_k\in {\rm MMPM}_{r_k}$, where each $r_k\geq 1$,
	$\sum_{k=1}^h r_k=r$, and each $\A_k$ is of the form (\ref{form:Ak}).
\end{Lemma}

\begin{IEEEproof}
	Induction on $r$. When $r=1$, then $h=1$, and the conclusion follows
	Lemma \ref{structure:0}. Assuming the conclusion holds for all
	$r<l$, when $r=l$, by Lemma \ref{structure:0} and applying the induction
	hypothesis to matrix $\B$, we get the conclusion.
\end{IEEEproof}

\begin{Lemma}\label{order:0}
	Let matrix $\A=(x^{u_i}\delta_{\tau(i),j})_{i,j\in [r]}$, where
	$\tau(i)=i-1\mod r$ for all $i\in [r]$, then for all $l\geq 1$,
	\be
	\boldsymbol{A}^l
	=\left(x^{\sum_{t\in [l]} u_{\tau^t(i)}}\delta_{\tau^l(i),j}\right)_{i,j\in [r]}.
	\label{Al:res}
	\ee
\end{Lemma}

\begin{IEEEproof}
	When $l=1$, the conclusion is trivial. Assume that the conclusion holds for all
	$l\leq k$. When $l=k+1$,
	\be\ba{lll}
	\boldsymbol{A}^{k+1}(i,j)
	&=& \ds \sum\limits_{v\in [r]}
	\boldsymbol{A}^k(i,v)\, \boldsymbol{A}(v,j) \\
	
	&=& \ds \sum\limits_{v\in [r]}
	x^{\sum_{t\in [k]} u_{\tau^t(i)}}\delta_{\tau^k(i),v}
	x^{u_v}\delta_{\tau(v),j}\\[2mm]
	
	&=&
	x^{\sum_{t\in [k+1]} u_{\tau^t(i)}}\delta_{\tau^{k+1}(i),j}.
	\ea
	\ee
\end{IEEEproof}

\begin{Lemma}\label{order:1}
	For the matrix $\A$ in Lemma \ref{order:0},
	\be
	\ord(\boldsymbol{A})=r\times
	\ord(x^{\sum_{i\in [r]} u_i}).
	\label{order:A}
	\ee
	As a corollary, for the matrix $\BP(1)$ in  Lemma \ref{structure:1},
	$\ord(\BP(1))={\rm LCM}({\rm ord}(\boldsymbol{A}_i), i=1..h)$.
\end{Lemma}

\begin{IEEEproof}
	By induction, it is easy to prove that $\tau^j(i)=i-j\mod r$ for all
	$i,j\in [r]$. So 
	$\{\tau^t(i)|t\in[r]\}=\{0,1,\cdots,r-1\}$. As a consequence,
	$x^{\sum_{t\in [r]}u_{\tau^t(i)}}=x^{\sum_{t\in [r]} u_t}$ is independent of $i$.
	
	By (\ref{Al:res}) and $\tau^r=1$,
	\be
	\boldsymbol{A}^r=x^{\sum_{i\in [r]} u_i}
	\boldsymbol{I}_{r\times r},
	\label{A:rtimes}
	\ee
	where
	$\boldsymbol{I}_{r\times r}$ is the identity matrix.
	So
	$
	\boldsymbol{A}^{r\times
		\ord(x^{\sum_{i\in [r]}u_i})}=\boldsymbol{I}_{r\times r}.
	$
	We prove that $m=r\times
	\ord(x^{\sum_{i\in [r]}u_i})$ is the smallest among all the positive integers $l$
	such that $\boldsymbol{A}^l=\boldsymbol{I}_{r\times r}$.
	
	Suppose there exists a number $1\leq c<m$ such that
	$\boldsymbol{A}^c=\boldsymbol{I}_{r\times r}$. Then
	$c=ar+b$ for some $a<\ord(x^{\sum_{i\in [r]}u_i})$ and $b\in[r]$.
	By (\ref{A:rtimes}),
	\be
	\boldsymbol{A}^c
	=(\boldsymbol{A}^r)^a\boldsymbol{A}^b
	=x^{a\sum_{i\in [r]} u_i} \boldsymbol{A}^b.
	\label{small:proof}
	\ee
	If $b\neq 0$, then $\tau^b\neq 1$, and
	by (\ref{Al:res}), $\boldsymbol{A}^b$ is not a diagonal matrix, so 
	$\boldsymbol{A}^c\neq \boldsymbol{I}_{r\times r}$ in (\ref{small:proof}).
	If $b=0$, then since $a<\ord(x^{\sum_{i\in [r]} u_i})$, 
	$\left(x^{\sum_{i\in [r]} u_i}\right)^a\neq 1$, again 
	$\boldsymbol{A}^c\neq \boldsymbol{I}_{r\times r}$ in (\ref{small:proof}).
\end{IEEEproof}

\begin{Example}~\label{Exa-use}
	Let
	\be
	\BP(1)=
	\begin{bmatrix}
		&x\\
		\I_{(r-1)\times (r-1)}&
	\end{bmatrix}.
	\label{ex:BP}
	\ee
	Then $\ord(\BP(1))=r\times \ord(x)=2Nr=q'$.
	When $r=1$, (\ref{ex:BP}) is just the monic monomial representation
	used in FHEW/TFHE. When setting $x=1$, (\ref{ex:BP}) is the
	the permutation matrix representation used in AP14.
	
	For any $c=ar+b$, where $a\in [2N],b\in [r]$,
	\be
	\BP(c)=
	\begin{bmatrix}
		0&x^{a+1}\boldsymbol{I}_{b\times b}\\
		x^a\boldsymbol{I}_{(r-b)\times (r-b)}&0\\
	\end{bmatrix}.
	\label{ex1:BPc}
	\ee
	In particular,	$\BP(r)=x\boldsymbol{I}_{r\times r}.$
	For any
	$\v=(v_i)_{i\in [r]}\in {\mathbb F}^r$,
	\begin{equation}
		\ba{lll}
		\BP(c)\v &=&
		(x^{a+1}v_b, x^{a+1}v_{b+1},\ldots,
		x^{a+1}v_{r-1}, \\
		
		&& \hfill
		x^a v_0, x^a v_1,\ldots, x^a v_{b-1}),
		\ea
		\label{equ-hadam}
	\end{equation}
	which equals the Hadamard product between vectors
	$(\underbrace{x^{a+1}, x^{a+1}, \ldots,x^{a+1}}_{r-b},
	\underbrace{x^a,x^a,\ldots,x^a}_{b})$ and
	$(v_b, v_{b+1}$, $\ldots, v_{r-1},v_0,v_1, \ldots,v_{b-1})$.
\end{Example}

Let $\mathbf{e}_1, \ldots, \mathbf{e}_r$ be the canonical basis of
${\mathbb F}^r$, namely, the $i$-th entry of $\mathbf{e}_i$ is 1, while all
other entries are 0. Set
\be
{\rm MMIV}_r:=\{x^i\mathbf{e}_j\,\big|\,i\in[2N],j\in[r]\}.
\ee

For any $\A=\left(x^{u_i}\delta_{i,\phi(j)}\right)_{i,j\in[r]}\in {\rm MMPM}_r$,
\be\ba{lll}
\A (x^k\mathbf{e}_l) &=&
\left(x^{u_i}\delta_{i,\phi(j)}\right)_{i,j\in[r]}
\left(x^k\delta_{l,m}\right)_{m\in [r]}\\

&=&
\left(x^{u_i+k}\delta_{i,\phi(l)}\right)_{i\in[r]} \\

&=& x^{u_{\phi(l)}+k}\mathbf{e}_{\phi(l)}.
\ea\ee
So ${\rm MMIV}_r$ is closed under the action of ${\rm MMPM}_r$.

\begin{Lemma}\label{orbits}
	Let $\BP(1)$ take the form in Lemma \ref{structure:1}, namely,
	$\BP(1)={\rm diag}(\A_1, \A_2, \ldots, \A_h)$, where for $1\leq k\leq h$,
	$\A_k\in {\rm MMPM}_{r_k}$ with $\sum_{k=1}^h r_k=r$, such that
	$\A_k(i,j)=x^{u_{k,i}}\delta_{j,i-1\ {\rm mod}\ r}$ for all $i,j\in [r_k]$.
	Then the action of group 
	$\BP({\mathbb Z}_{2Nr})$ on set ${\rm MMIV}_r$ is regular, and
	the number of orbits is
	\be
	\sum\limits_{i=1}^h \frac{2Nr_i}{\ord(\boldsymbol{A}_i)}.
	\label{number:orbits}
	\ee
\end{Lemma}

\begin{IEEEproof} Since ${\rm MMIV}_r$ contains the basis $\{\e_j\,|\,j\in[r]\}$ of
	${\mathbb F}^r$, any linear transformation of ${\mathbb F}^r$ leaving each
	element of ${\rm MMIV}_r$ invariant must be the identity transformation. So 
	the action of $\BP({\mathbb Z}_{2Nr})$ on ${\rm MMIV}_r$ is regular.
	Below we prove (\ref{number:orbits}) by induction.
	
	Set
	\be
	\A={\rm diag}(\A_1, \I_{r_2\times r_2}, \ldots, \I_{r_h\times r_h}).
	\label{proof:A1}
	\ee
	Then $\A$ generates a subgroup $\langle \A \rangle$ of $MP({\mathbb F}^r)$.
	We prove that the number of orbits in $\cal X$ under the action of
	$\langle \A \rangle$ is $2Nr_1/\ord(\A_1)$.
	
	By (\ref{order:A}),
	\be
	\ord(\boldsymbol{A}_1)=r_1\times \ord(x^{v_0}),
	\hbox{ where } v_0=\sum_{t\in [r_1]} u_{1,t}.
	\ee
	Now that $x$ generates a cyclic group $\langle x\rangle$ of order $2N$, and
	$x^{v_0}$ generates a subgroup $\langle x^{v_0}\rangle$ of order
	$\ord(x^{v_0})=\ord(\boldsymbol{A}_1)/r_1$, there are
	\be
	c:=2N/\ord(x^{v_0})=2Nr_1/\ord(\boldsymbol{A}_1)
	\label{def:c}
	\ee
	cosets of
	$\langle x^{v_0}\rangle$ in $\langle x\rangle$. Select one element from
	each coset respectively, and let them be $x^{v_0}, x^{v_1}, \ldots,
	x^{v_{c-1}}$.
	
	We claim (1) under the action of subgroup $\langle \A\rangle$, for any
	$i,j\in [c]$ such that $i\neq j$, the orbits of $x^{v_i}\e_0$ and
	$x^{v_j}\e_0$ do not overlap. Suppose the converse is true, namely
	there exists $d\in [\ord(\boldsymbol{A})]$ such that
	$\A^d x^{v_i}\e_0=x^{v_j}\e_0$. As $r_1|\ord(\boldsymbol{A})$, let
	$d=ar_1+b$, where $b\in [r_1]$. By (\ref{A:rtimes}),
	$\boldsymbol{A}^{r_1}=x^{v_0}\boldsymbol{I}_{r\times r}$, so
	\be
	\A^d x^{v_i}\e_0=\A^b (\A^{r_1})^a x^{v_i}\e_0
	=x^{av_0+v_i} \A^b\e_0
	=x^{v_j}\e_0.
	\label{proof:orbit}
	\ee
	
	If $b\neq 0$, by (\ref{Al:res}), $\A^b$ does not preserve the 1-space
	spanned by $\e_0$, and the last equality in (\ref{proof:orbit}) is false. So $b=0$,
	and $x^{av_0}x^{v_i}=x^{v_j}$. This indicates that $x^{v_i}, x^{v_j}$
	are in the same coset of $\langle x^{v_0}\rangle$ in $\langle x\rangle$,
	contradiction. This proves claim (1).
	
	We claim (2) for any $i\in [c]$, the number of elements in the orbit
	of $x^{v_i}\e_0\in {\cal X}$ under $\A$ is at least
	$\ord(\boldsymbol{A})$.
	To prove the claim we only need to show that for
	any $s,t\in [\ord(\boldsymbol{A})]$ such that $s\neq t$,
	$\A^s x^{v_i}\e_0\neq \A^t x^{v_i}\e_0$.
	
	Suppose the converse is true. Without loss of generality, assume $s>t$.
	Then $0<s-t<\ord(\boldsymbol{A})$, and $\A^{s-t} \e_0=\e_0$. On the other hand,
	by (\ref{Al:res}), $\A^{s-t}$ does not preserve the 1-space
	spanned by $\e_0$, contradiction. This proves claim (2).
	
	Set
	\be
	{\rm MMIV}_{r_1}:=\{x^i\mathbf{e}_j\,\big|\,i\in[2N],j\in[r_1]\}.
	\ee
	Obviously every element of ${\rm MMIV}_{r}\backslash {\rm MMIV}_{r_1}$ is fixed
	by $\A$, and the set ${\rm MMIV}_{r_1}$ is invariant under $\A$.
	
	We claim (3) ${\rm MMIV}_{r_1}$ is the union of the orbits of
	$\{x^{v_i}\e_0, i\in [c]\}$ under the action of $\langle \A\rangle$.
	Denote by ${\rm orbit}(x^{v_i}\e_0)$ the orbit of $x^{v_i}\e_0$ under $\A$,
	and for any set $S$, denote by $\#S$ the number of elements in the set.
	By claims (1), (2), and using (\ref{def:c}), we get
	\be\ba{lll}
	2Nr_1=\#{\rm MMIV}_{r_1}
	&\geq& \ds \sum_{i\in [c]} \#{\rm orbit}(x^{v_i}\e_0)\\
	
	&\geq& \ds \sum_{i\in [c]} \ord(\boldsymbol{A})
	= 2Nr_1.
	\ea
	\ee
	So ${\rm MMIV}_{r_1}= \cup_{i\in [c]} {\rm orbit}(x^{v_i}\e_0)$.
	This proves claim (3). As a corollary, $\#{\rm orbit}(x^{v_i}\e_0)=\ord(\boldsymbol{A})$
	for all $i\in [c]$, and the number of orbits in ${\rm MMIV}_{r}$ under
	the action of $\langle \A\rangle$ is $c$.
	
	When $\A$ takes the general form
	${\rm diag}(\A_1, \A_2, \ldots, \A_h)$, by induction, the conclusion (\ref{number:orbits})
	can be easily proved.
\end{IEEEproof}

\begin{theorem}\label{ultimate}
	For any group representation
	$\BP: {\mathbb Z}_{2Nr}\longrightarrow {\rm MMPM}_{r}$,
	the action of $\BP({\mathbb Z}_{2Nr})$ on
	${\rm MMIV}_{r}$ is transitive if and only if
	$\BP(1)$ is similar to
	\be
	\begin{bmatrix}
		&&&&x^{u_0}\\
		x^{u_1}&&&&\\
		&x^{u_2}&&&\\
		&&\ddots&&\\
		&&&x^{u_{r-1}}&
	\end{bmatrix},
	\label{BP1:form}
	\ee
	where $\ord(x^{\sum_{i\in [r]} u_i})=2N$.
\end{theorem}

\begin{IEEEproof}
	By (\ref{number:orbits}) and the fact that
	$\ord(\boldsymbol{A}_i)|2Nr_i$ for all $1\leq i\leq h$,
	in order for the number of orbits to be 1, it is both sufficient and
	necessary that $h=1$ and $2Nr_1=\ord(\boldsymbol{A}_1)$. When $h=1$,
	by (\ref{order:A}),
	the latter condition can be written as $2Nr=\ord(\BP(1))=r\times
	\ord(x^{\sum_{i\in [r]} u_i})$.
\end{IEEEproof}

Only cyclic subgroups of order $2Nr$ and transitive on ${\rm MMIV}_{r}$
can have test-coefficient look-up property,
because for any test polynomial vector $\v\in {\mathbb F}^r$, 
for any monomial $\lambda_ix^i$ in the $j$-th entry of $\v$, 
only in such a subgroup can there be
one and only one element that changes the monomial to
$\lambda_ix^0=\lambda_i$, {\it i.e.}, the constant term, 
in the first entry of the resulting vector in ${\mathbb F}^r$.
In the following, we consider the design of test polynomial vectors
encoding nega-cyclic functions.

\begin{Lemma}
	\label{lem:vec}
	Let $\BP: {\mathbb Z}_{2Nr}\longrightarrow {\rm MMPM}_{r}$ be a group
	representation taking the form of (\ref{BP1:form}),
	where $\ord(x^{\sum_{i\in [r]} u_i})=2N$.
	Then
	$
	\BP(Nr)=-\I_{r\times r},
	$
	and for any $k\in [2Nr]$, the elements in set
	\be
	\{\BP(k+i)\mathbf{e}_0\,\big|\,i\in[Nr]\}\subset {\rm MMIV}_{r}
	\label{BP:independent}
	\ee
	are $\mathbb Q$-linearly independent vectors.
\end{Lemma}

\begin{IEEEproof} The first conclusion is direct from (\ref{A:rtimes})
	and $\ord(x^{\sum_{i\in [r]} u_i})=2N$.
	For the second conclusion,
	for any $1\leq l\leq Nr$,
	set
	\be
	{\cal Y}_l:=\{\BP(k+i)\mathbf{e}_0, \ i\in [l]\}.
	\ee
	We prove that every ${\cal Y}_l$ is a $\mathbb Q$-linearly independent set of vectors.
	
	When $l=1$, ${\cal Y}_l=\{\BP(k)\mathbf{e}_0\}$ is
	$\mathbb Q$-linearly independent.
	Assume that for all $l\leq h<Nr$, ${\cal Y}_h$ is
	$\mathbb Q$-linearly independent. For $l=h+1$, if ${\cal Y}_{h+1}$ is
	not $\mathbb Q$-linearly independent, then
	$\BP(k+h)\mathbf{e}_0=\sum_{i\in [h]} v_i \BP(k+i)\mathbf{e}_0$
	for some $v_i\in {\mathbb Q}$. 
	
	In ${\rm MMIV}_{r}$, it must be that
	$\BP(k+h)\mathbf{e}_0=-\BP(k+i)\mathbf{e}_0$ for some $i\in [h]$, namely,
	$\BP(h-i)\mathbf{e}_0=-\mathbf{e}_0$ for some $0<h-i<Nr$.
	By (\ref{Al:res}), $h-i=mr$ for some $0<m<N$. By
	(\ref{A:rtimes}), $x^{m\sum_{i\in [r]} u_i}=-1$.
	So $x^{2m\sum_{i\in [r]} u_i}=1$, contradicting with
	$\ord(x^{\sum_{i\in [r]} u_i})=2N$.
\end{IEEEproof}

\vskip .2cm
Let $(v_0, v_1, \ldots, v_{2Nr-1})^T$ be any vector in
${\mathbb Z}_{2Nr}$. For any $k\in [2Nr]$, set
\be
\v_k:=\sum_{i\in [Nr]} v_i \BP(k-i) \e_0\in {\mathbb F}^r.
\label{def:v}
\ee
By Lemma \ref{lem:vec}, $\{\BP(k-i) \e_0\,\big|\,i\in [Nr]\}$
is $\mathbb Q$-linearly independent. It is easy to see that 
for any $k\in [2Nr]$, $\v_{k+ Nr \ {\rm mod}\ 2Nr}=-\v_k$, and 
for any $l\in [2Nr]$,
$\BP(l)\v_k=\v_{k+l \ {\rm mod}\ 2Nr}$.

The following lemma indicates that $\v_0$ can serve 
as the test polynomial vector.

\begin{Lemma} [Test-coefficient look-up property]
	\label{lem:123}
	For the group representation $\BP:
	{\mathbb Z}_{2Nr}\longrightarrow {\rm MMPM}_{r}$ in Lemma
	\ref{lem:vec}, and for $\v_0$ defined in (\ref{def:v}) where $k=0$, 
	if $v_{Nr+l}=-v_l$ for all $l\in [Nr]$, then 
	for any $i\in [2Nr]$, 
	$\BP(i)$ changes 
	the term $v_i\BP(-i)\e_0$ of $\v_0$ into the term $v_i\e_0$
	of $\v_i$. 
\end{Lemma}

\begin{IEEEproof}
	No matter if $i\in [Nr]$ or not, 
	it is always true that $v_i\BP(-i)\e_0$ is a term of $\v_0$, 
	and 
	\be
	\BP(i)(v_i\BP(-i)\e_0)=v_i\e_0=v_i\BP(i-i)\e_0
	\ee 
	is a term of $\v_i$, {\it i.e.}, 
	the constant-term in the first entry of $\v_i$. 
\end{IEEEproof}

\begin{Example}\label{Exa-ult}
	For $\BP$ taking the form of Lemma \ref{lem:vec}, 
	let $-i=ar+b\in [2Nr]$, where $a\in[N], b\in[r]$. 
	By (\ref{Al:res}), the $(b,0)$-entry of matrix $\BP(b)$ is 
	$x^{\sum_{t\in[b]}u_{b-t}}$, 
	while all other entries in the first column of $\BP(b)$ are zero, namely,  
	$\BP(b)\mathbf{e}_0=x^{\sum_{t\in[b]}u_{b-t}}\mathbf{e}_b$.
	By (\ref{A:rtimes}), $\BP(r)=x^{\sum_{t\in[r]} u_t}\I_{r\times r}$. 
	So
	\begin{equation}
		\BP(-i)\mathbf{e}_0=\BP(r)^a\BP(b)\mathbf{e}_0
		=x^{a\sum_{t\in[r]} u_t+\sum_{t\in[b]} u_{b-t}}\mathbf{e}_b,
		\label{equ:basis-test-gene}
	\end{equation}
	and
	\begin{equation}
		\begin{split}
			\boldsymbol{v}_0&=\sum\limits_{i\in[Nr]}v_i\BP(-i)\mathbf{e}_0\\
			&=\sum\limits_{a\in[N],b\in[r]}v_{-(ar+b)}
			x^{a \sum_{t\in[r]} u_t +\sum_{t\in[b]} u_{b-t}}\mathbf{e}_b.
		\end{split}
		\label{equ:test-gene}
	\end{equation}
	
	In particular, if $\BP$ takes the form (\ref{ex:BP}), then for any nega-cyclic
	function $f: {\mathbb Z}_{2Nr}\longrightarrow {\mathbb Q}$,
	set $v_i=f(i)$ for all $i\in [2Nr]$. Then
	$v_{Nr+k}=-v_k$ for all $k\in [Nr]$.
	The test polynomial vector $\v_{\rm test}=\v_0$ is
	\be\label{equ:test-use}
	\hskip -.7cm
	\ba{ll}
	& \ds\sum\limits_{b\in[r],\, a\in [N]} f(-(ar+b))x^a\mathbf{e}_b
	\\
	
	=& \ds \hskip -.2cm
	\begin{pmatrix}
		\scriptstyle	f(0)+f(-r)x+\cdots+f(-cr)x^c+\cdots+f(-(N-1)r)x^{N-1}\\
		\scriptstyle	f(-1)+f(-1-r)x+\cdots+f(-1-cr)x^c+\cdots+f(-1-(N-1)r)x^{N-1}\\
		\vdots\\
		\scriptstyle	f(-d)+f(-d-r)x+\cdots+f(-d-cr)x^c+\cdots+f(-d-(N-1)r)x^{N-1}\\		
		\vdots\\
		\scriptstyle	f(1-r)+f(1-2r)x+\cdots+f(1-(c+1)r)x^c+\cdots+f(1-Nr)x^{N-1}\\
	\end{pmatrix}.
	\ea
	\ee
	
	For any $m=-d-cr\in [2Nr]$, where $c\in [2N], d\in [r]$, 
	$\BP(m)\v_{\rm test}$ equals
	\be\label{equ:ex}
	\hskip -.3cm
	\ba{ll}
	& \ds\sum\limits_{b\in[r],\, a\in [N]} f(-(ar+b))x^{a-c}\mathbf{e}_{b-d}
	\\
	
	=& \ds \hskip -.2cm
	\begin{pmatrix}
		\scriptstyle	f(m)+f(m-r)x+\cdots+f(m-(N-1)r)x^{N-1}\\
		\scriptstyle	f(m-1)+f(m-r-1)x+\cdots+f(m-(N-1)r-1)x^{N-1}\\
		\vdots\\
		\scriptstyle	f(m-r+1)+f(m-2r+1)x+\cdots+f(m-Nr+1)x^{N-1}\\
	\end{pmatrix}.
	\ea
	\ee
	It is easy to see that for any $m =-cr-d\in[q']$, $\BP(m)$ changes the term $f(m)\BP(-m)\mathbf{e}_0=f(m)x^c\mathbf{e}_d$ of $\v_0$,\emph{i.e.}, the $(c+1)$-st term in the $(d+1)$-st row of test polynomial vector $\v_0$, into the term $f(m)\e_0$ of $\v_m$, \emph{i.e.}, the constant term in the first row of the resulting polynomial vector. This is the test-coefficient look-up property demanded in bootstrapping.
\end{Example}

\section{New Functional Bootstrapping Scheme}
From now on, we always set $\BP(1)$ to be of the form (\ref{ex:BP}).
By Theorem \ref{ultimate}, this choice does not lose generality.

Given a plaintext vector $\m\in {\cal R}_t^r$, its RLWE
encryption is done component-wise, resulting in a vector of RLWE
ciphertexts, called an $r$-dimensional {\it RLWE vector} encrypting $\m$.
The space of $r$-dimensional ${\rm RLWE}_{N,Q}$ vectors is
denoted by ${\rm RLWE}_{N,Q}^r$. Any $r$-dimensional ${\rm RLWE}_{N,Q}$ vector is a matrix
in ${\rm Mat}_{r\times 2}({\cal R_{N,Q}})$.

According to (\ref{equ-hadam}), for any $c\in {\mathbb Z}_{2Nr}$, for any
any $r$-dimensional ${\rm RLWE}_{N,Q}$ vector $\v$
encrypting a plaintext vector $\m$, the usual matrix product of $r\times r$ plaintext
matrix $\BP(c)$ with $r\times 2$ matrix $\v$,
results in an ${\rm RLWE}_{N,Q}$ vector ($r\times 2$ matrix)
encrypting plaintext vector $\BP(c)\m$. This defines the multiplication between
plaintext matrix $\BP(c)$ and RLWE vector $\v$. 

Let $\s=(s_i)_{i=1..n}\in {\mathbb Z}_3^n$, and let 
$s_{i+}=\max(s_i,0)$ and $s_{i-}=\max(-s_i,0)$.
If the $s_{i+}, s_{i-}$ are each encrypted as an RGSW ciphertext,
then the ciphertext multiplication between RLWE vector $\v$ and
the RGSW ciphertext encrypting $s_{i+}$ (or $s_{i-}$) is done
component-wise between each component of the RLWE vector and the
the RGSW ciphertext. The result is an RLWE vector encrypting
$s_{i+}\m$ (or $s_{i-}\m$). This defines the multiplication between
an RLWE vector and an RGSW ciphertext. 

By (\ref{equ-exten-cmux}), for any $\m\in {\mathbb F}^r$,
\be\ba{ll}
& \CMux(s_{i+}, s_{i-};\BP(-a_i), \I_{r\times r}, \BP(a_i))\ \m\\

=& \m+(\BP(-a_i)\m-\m)s_{i+}+(\BP(a_i)\m-\m)s_{i-}.
\ea
\label{cmux:BP}
\ee
The expression can be evaluated homomorphically if $\m, 
s_{i+}, s_{i-}$ are encrypted as RLWE vector, RGSW ciphertext,
RGSW ciphertext respectively, and the multiplications
are between an RLWE vector and an RGSW ciphertext. 

We are ready to extend the classical TFHE scheme to a new scheme ``\texttt{BootMMPM}"
based on the monic monomial permutation matrix representation $\BP$ of
${\mathbb Z}_{2Nr}$ where $\BP(1)$ takes the form (\ref{ex:BP}), as follows:

\begin{figure}[!t]
	\begin{algorithm}[H]
		
		\caption{$\texttt{BootMMPM}(\texttt{ct},f)$, where $\texttt{ct}$ is an 
			${\rm LWE}_{n,q}$ ciphertext with ternary secret, 
			$f: {\mathbb Z}_t\longrightarrow {\mathbb Z}_{t'}$ is nega-cyclic.
		}
		\label{alg:LSB1}
		\begin{algorithmic}[1]
			
			\REQUIRE 
			\begin{itemize}
				\item $\texttt{ct}\in {\rm LWE}_{n,q}(m,\s)$ with secret 
				$\s=(s_k)_{k=1..n}\in \{1,0,-1\}^n$, plaintext $m\in {\mathbb Z}_t$;
				
				\item test polynomial vector $\v_{\rm test}\in {\mathbb Z}_{t'}$ whose coefficients
				encode a nega-cyclic extension 
				$f': {\mathbb Z}_{2Nr}\longrightarrow {\mathbb Z}_{t'}$ of $f$; 
				
				\item bootstrapping keys: for all $k=1, \ldots, n$, 
				$\texttt{ct}_{k+}, \texttt{ct}_{k-}$ are ${\rm RGSW}_{N,Q}$ ciphertexts encrypting
				$s_{k+}, s_{k-}$ respectively with secret $z\in {\cal R}_{N,Q}$, the gadget base is
				$B$, and $l_B=\lceil \log_B Q\rceil$;
				
				\item key-switching keys 
				from ${\rm LWE}_{N,Q}$ ciphertexts with secret $\overrightarrow{z}$
				to ${\rm LWE}_{n,Q}$ ciphertexts with secret $\s$, the gadget base is 
				$B_{\rm KS}$, and $l_{\rm KS}=\lceil \log_{B_{\rm KS}} Q\rceil$.
			\end{itemize}
			
			\ENSURE $\texttt{ct}\in {\rm LWE}_{n,q}(f(m),\s)$.
			
			\STATE [Modulus switch]
			$\texttt{ct}=(a_1, a_2, \ldots, a_n,b)\in {\mathbb Z}_{2Nr}^{n+1}
			\leftarrow {\rm ModulusSwitch}_{q\rightarrow 2Nr}(\texttt{ct})$.
			
			\STATE [Phase accumulation]
			$ACC\leftarrow \BP(b)(0,\v_{\rm test}\times \lfloor Q/t'\rfloor)$
			
			\FOR{$k=1$ to $n$}
			\STATE \ \ $ACC\leftarrow {\rm CMux}(\texttt{ct}_{k+},\texttt{ct}_{k-}; 
			\BP(-a_k),\I_{r\times r},\BP(a_k))$\\
			\hfill $\times ACC$
			\ENDFOR
			
			\STATE [Constant-term extraction]
			
			$\texttt{ct}\leftarrow$ SampleExtract (1st entry of vector $ACC$)
			
			\STATE [Key switch]
			$\texttt{ct}\leftarrow {\rm KeySwitch}_{\overrightarrow{z}\rightarrow 
				\boldsymbol{s}}(\texttt{ct})$
			
			\STATE [Modulus switch]
			$\texttt{ct}\leftarrow {\rm ModulusSwitch}_{Q\rightarrow q}(\texttt{ct})$
			
		\end{algorithmic}
	\end{algorithm}
\end{figure}

The algorithm would be correct if we could control the error bound in the output ciphertext,
and in particular, if the error growth in the homomorphic CMux operation is slow.
In the product of a plaintext matrix $\BP(c)$ and an RLWE vector $\v$, the error
bound is the same with that of $\v$, due to the Hadamard product nature of this product.

The $B$-digit decomposition of an 
RLWE vector is done component-wise. Recall that when an RLWE ciphertext
takes the form of a 2-dimensional row vector, after digit decomposition,
it becomes an $2l_B$-dimensional row vector. So
an $r$-dimensional ${\rm RLWE}_{N,Q}$ vector as a matrix of size $r\times 2$, after
digit decomposition, becomes a matrix of size $r\times 2l_B$.

Recall that
an RGSW ciphertext is a matrix in ${\rm Mat}_{2l_B\times 2}({\cal R}_{N,Q})$.
The multiplication of an RLWE vector $\v$ with an RGSW ciphertext $\C$
encrypting an element of $\{1,0\}$ is done component-wise, resulting in
a new RLWE vector $\v'=\left(\g_B^{-1}(\v)\right)\C$, where
$\g_B^{-1}$ is the $B$-digit decomposition operation. 

Let the error vectors in $\v, \v', \C$ be $\e\in {\cal R}_{N,Q}^r$,
$\e'\in {\cal R}_{N,Q}^r$, $\e_C\in {\cal R}_{N,Q}^{2l_B}$ respectively, then 
as in the case of $r=1$, we have
\be
\e'=(\g_B^{-1}(\v))\e_C+\e.
\label{error:grow}
\ee
By Corollary 3.15 of \cite{chillotti2016faster}, if
$\e, \e_C$ are subgaussian with variance proxy 
$\beta, \sigma$ respectively, then 
$\e'$ is subgaussian with variance proxy
\be
2Nl_B B^2\sigma^2+\beta^2.
\label{GSW:error}
\ee

\begin{Lemma}
	Let $d\in \{1,0,-1\}$, and let $d_{+}, d_{-}$ be defined as in (\ref{d:pm}).
	Let $\c_1, \c_0, \c_{-1}$ be three ${\rm RLWE}_{N,Q}$ vectors where the errors
	are all subgaussian with variance proxy $\beta^2$, and let
	$s_+, s_-$ be ${\rm RGSW}_{N,Q}$ ciphertexts encrypting
	$d_{+}, d_{-}$ respectively,
	where the errors are independent subgaussian with variance proxy $\sigma^2$.
	If the errors in $\c_1, \c_0, \c_{-1}$ are independent of
	the errors in $d_{+}, d_{-}$,
	then the homomorphic evaluation result of $\CMux(d_+,d_-;\c_1, \c_0, \c_{-1})$
	has a subgaussian error vector $\e_{\rm CMux}\in {\cal R}_{N,Q}^r$ with variance proxy
	\be
	\beta^2+4Nl_B B^2\sigma^2.
	\ee
	\label{cmux}
\end{Lemma}

\begin{IEEEproof}
	Let the error vectors in $d_{+}, d_{-}, \c_1, \c_0, \c_{-1}$ be
	$\e_{+}, \e_{-}, \e_1, \e_0, \e_{-1}$ respectively, where
	the first two are in ${\cal R}_{N,Q}^{2l_B}$, the latter three are in
	${\cal R}_{N,Q}^r$. 
	For a subgaussian random variable $v$, we use $\Var(v)$ to denote its
	variance proxy.
	
	By (\ref{equ-exten-cmux}) and (\ref{error:grow}),
	\be
	\begin{split}
		\e_{\rm CMux}=\e_0+\left(\boldsymbol{g}^{-1}
		(\boldsymbol{c}_1-\boldsymbol{c}_0)\right)\boldsymbol{e}_+
		+(\boldsymbol{e}_1-\boldsymbol{e}_0)s_+
		\\
		+\left(\boldsymbol{g}^{-1}(\boldsymbol{c}_{-1}-\boldsymbol{c}_0)\right)
		\boldsymbol{e}_-+(\boldsymbol{e}_{-1}-\boldsymbol{e}_0)s_-.
	\end{split}
	\label{cmux:e}
	\ee
	Since $
	\boldsymbol{e}_0+(\boldsymbol{e}_1
	-\boldsymbol{e}_0)s_++(\boldsymbol{e}_{-1}-\boldsymbol{e}_0)s_-
	\in
	\{\boldsymbol{e}_0,\boldsymbol{e}_1,\boldsymbol{e}_{-1}\},
	$
	we have
	${\rm Var}(\boldsymbol{e}_0+(\boldsymbol{e}_1-\boldsymbol{e}_0)s_+
	+(\boldsymbol{e}_{-1}-\boldsymbol{e}_0)s_-)
	=\beta^2$.
	
	By the error independence assumption and (\ref{GSW:error}),
	we get
	\be
	\ba{lll}
	\Var(\e_{\rm CMux})
	&=&
	\Var(\boldsymbol{e}_0+(\boldsymbol{e}_1-\boldsymbol{e}_0)s_+
	+(\boldsymbol{e}_{-1}-\boldsymbol{e}_0)s_-)
	\\
	&& \hfill +\Var(\boldsymbol{g}^{-1}
	(\boldsymbol{c}_1-\boldsymbol{c}_0)\boldsymbol{e}_+)\phantom{m}\\
	&& \hfill 
	+\Var(\boldsymbol{g}^{-1}
	(\boldsymbol{c}_{-1}-\boldsymbol{c}_0)\boldsymbol{e}_-)
	\\
	&=& \beta^2
	+4Nl_B B^2\sigma^2.
	\ea
	\ee
\end{IEEEproof}

\begin{theorem}
	In the input of Algorithm~\ref{alg:LSB1}, let
	$\sigma^2, \sigma^2_{\rm KS}$ be respectively the variance proxies
	of the bootstrapping keys and key-switching keys. 
	Then for any input ${\rm LWE}_{n,q}$ ciphertext encrypting a message $m$,
	as long as after modulus switch from $q$ to $2Nr$, the ciphertext is
	correctly decryptable, 
	Algorithm~\ref{alg:LSB1} outputs an ${\rm LWE}_{n,q}$ ciphertext
	encrypting message $f(m)$ with a subgaussian error of variance proxy 
	\be
	4nNl_BB^2(q/Q)^2\sigma^2+Nl_{\rm KS}(q/Q)^2\sigma_{\rm KS}^2
	+(\|\boldsymbol{s}\|_2^2+1)/12.
	\ee
\end{theorem}

\begin{IEEEproof}
	In computing $ACC$, the initial value $\BP(b)(0,\boldsymbol{v}_{\rm test}\times \lfloor Q/t'\rfloor)$ 
	is error-free. By Lemma \ref{cmux},
	every round of the for-loop increases the variance proxy of the
	subgaussian error by $4Nl_BB^2\sigma^2$. So after the phase accumulation loop,
	the error has variance proxy $4nNl_BB^2\sigma^2$.
	
	The constant-term extraction of the first row of $ACC$ 
	changes the RLWE ciphertext with secret $z$ into an LWE ciphertext
	with secret $\overrightarrow{z}$, but does not change the
	the variance proxy of the error.
	
	In the key switch procedure,  
	by Lemma 6 of \cite{ducas2015fhew}, for any ${\rm LWE}_{N,Q}$ ciphertext
	with subgaussian error of variance proxy $\beta^2$, the 
	procedure generates an ${\rm LWE}_{n,Q}$ ciphertext with 
	subgaussian error of variance proxy 
	$\beta^2+Nl_{\rm KS}\sigma_{\rm KS}^2$. 
	
	In the modulus switch procedure in the last step of the algorithm,
	by \cite{ducas2015fhew}, for any ${\rm LWE}_{n,Q}$ ciphertext
	with subgaussian error of variance proxy $\beta^2$, the modulus switch
	from $Q$ to $q$ based on non-randomized rounding
	function $\lfloor\cdot\rceil$, generates an ${\rm LWE}_{n,q}$ ciphertext with 
	subgaussian error of variance proxy 
	\be
	(q/Q)^2 \beta^2+ (\|\boldsymbol{s}\|_2^2+1)/12,
	\label{err:mod}
	\ee
	where $\|\boldsymbol{s}\|_2$ is the $L_2$-norm of the secret $\s$, and 
	$1/12=(0.5-(-0.5))^2/12$ is the variance of
	the uniform distribution on $[-1/2,1/2)$. 
	So the modulus switch
	modifies the variance proxy of
	the subgaussian error by first scaling it with $(q/Q)^2$, then adding a term 
	of rounding error $(\|\boldsymbol{s}\|_2^2+1)/12$. 
\end{IEEEproof}

\vskip .2cm
For general functional bootstrapping where the evaluation function 
$f: {\mathbb Z}_t\longrightarrow {\mathbb Z}_{t'}$ is not nega-cyclic, 
the routine procedure used in FHEW/TFHE \cite{chillotti2020tfhe, liu2021large}
is to take the input plaintext $m$ as an integer in
$[0,t)$, take the plaintext modulus as $2t$, and take
$f$ as a {\it cyclic function} defined on ${\mathbb Z}_{2t}$, namely, $f(x+t)=f(t)$ for all 
$x\in [-t,t)$. The plaintext encrypted in the input 
ciphertext, now being in ${\mathbb Z}_{2t}$, has an extra bit as the new MSB or sign bit,
whose value is unknown. 

On the other hand,
the original plaintext $m\in [0,t)$ is positive, so the new MSB must be removed before
functional bootstrapping based on $f$.
To extract the new MSB, one bootstrapping based on nega-cyclic function 
\be
{\rm mp}(x):=\left\{\ba{ll}
-t/2, & \hbox {if } x\in [-t/2,t/2), \\
t/2, & \hbox {if } x\in [-t,t)\backslash [-t/2,t/2)
\ea\right.
\label{rel:mp}
\ee
is sufficient, because for any $x\in [-t,t)$,
\be
{\rm sign}(x)\times t={\rm mp}(x)+t/2.
\label{rel:sign}
\ee

After being extracted, the MSB is subtracted from the 
input plaintext, so that after this modification, the plaintext is always in $[0,t)$. Then another
bootstrapping based on the trivial nega-cyclic extension of $f$ from $[0,t)$ to $[-t,t)$
suffices.

\section{Complexity and experiments}
In \texttt{BootMMPM}, the most important factor is the dimension $r$
of the test polynomial vector. By (\ref{init:r}), 
$r\geq q/(2N)$, and $r$ should be as small as possible. 

On the other hand, the choice of $r$ should guarantee that 
the ciphertext after modulus switch from $q$ to $q'=2Nr$ is decryptable.
Let the error of the input ciphertext be subgaussian with
variance proxy $\beta^2$. 
By (\ref{err:mod}), the modulus switch from $q$ to $q'$ changes the 
variance proxy to
\be
\sigma^2:=(2Nr/q)^2 \beta^2+ (\|\boldsymbol{s}\|_2^2+1)/12.
\label{def:s2}
\ee
For the ciphertext after modulus switch to be decryptable,
it is sufficient that the heuristic bound of the error $H\sigma<2Nr/(2t)$, 
where $H=O(1)$ is a constant determined by the decryption failure probability.
By (\ref{def:s2}), the inequality becomes
\be
N^2r^2>\frac{\|\boldsymbol{s}\|_2^2+1}{12}\Big/
\left(\frac{1}{H^2t^2}-\frac{4\beta^2}{q^2}
\right).
\label{bound:r}
\ee

When 
\be\ba{lll}
N=\tilde{O}(n), &
\|\boldsymbol{s}\|_2^2=\tilde{O}(n),& 
\beta=\tilde{O}(n^{d_e}), \\

r=\tilde{O}(n^{d_r}), &
q=\tilde{O}(n^{d_q}), & t=\tilde{O}(n^{d_t}),
\ea 
\label{set:par}
\ee
where the $d_i$ are positive real numbers, $d_q\geq d_t+d_e$ 
and $d_e\geq 1/2$, 
then (\ref{bound:r}) requires $d_r\geq d_t$. On the other hand, 
$r\geq q/(2N)$ requires $d_r\geq d_q-1\geq d_t+d_e-1$.
So
\be
d_r\geq \max(d_t, d_q-1).
\ee

The efficiency factors of Algorithm \texttt{BootMMPM} on the server's side include memory cost
and run-time cost. The former is heavily influenced by the size of bootstrapping keys and
key-switching keys, while 
the latter mainly depends on the time complexity of the phase accumulation loop.

(1) Bootstrapping key size. 

There are $2n$ bootstrapping keys, and each key is an RGSW ciphertext that is a matrix 
in ${\rm Mat}_{2l_B\times 2}({\cal R}_{N,Q})$. Every element of ${\cal R}_{N,Q}$ is a polynomial 
of degree $<N$ with coefficient in ${\mathbb Z}_Q$, so it has 
$N\log Q$ bits. Overall, in \texttt{BootMMPM}
the total number of bits in the bootstrapping keys is
$8nl_BN\log Q$.

In contrast, directly using TFHE to make functional bootstrapping requires
$r$ to be a power of 2, and the ring to be ${\cal R}_{Nr,Q}$.
The bootstrapping keys in TFHE have  
$(8nl_BN\log Q)r$ bits. In the setting of (\ref{set:par}), this size is
polynomially bigger than that in \texttt{BootMMPM}.

(2) key-switching key size. 

The key-switching keys are the
${\rm LWE}_{n,Q}(vz_iB_{\rm KS}^j)$ for all $v\in [B_{\rm KS}], i\in[N], j\in[l_{\rm KS}]$.
So in \texttt{BootMMPM}, 
the total number of bits in the key-switching keys is 
$NB_{\rm KS}l_{\rm KS}(n+1)\log Q$ bits. 

In contrast, in TFHE the ring is ${\cal R}_{Nr,Q}$, so 
the key-switching keys in TFHE have $(NB_{\rm KS}l_{\rm KS}(n+1)\log Q)r$ bits.
In the setting of (\ref{set:par}), this size is
polynomially bigger than that in \texttt{BootMMPM}..

(3) Phase-accumulation time complexity.

The phase accumulation procedure consists of 
$n$ CMux operations. In \texttt{BootMMPM}, each CMux requires 2 multiplications between
an RLWE vector and an RGSW ciphertext, or equivalently, $2r$ 
multiplications between an RLWE ciphertext and an RGSW ciphertext.

The multiplication between an RLWE ciphertext and an RGSW ciphertext
is the usual matrix product between a matrix in ${\rm Mat}_{2\times 2l_B}({\cal R}_{N,Q})$ and
a matrix in ${\rm Mat}_{2l_B\times 2l_B}({\cal R}_{N,Q})$. It contains
$4l_B$ polynomial multiplications in ${\cal R}_{N,Q}$. By FFT, each 
polynomial multiplication requires at most $(3/2)N\log N$ integer 
multiplications in ${\mathbb Z}_Q$.
So the multiplication between an RLWE ciphertext and an RGSW ciphertext has
integer multiplication complexity $6l_BN\log N$. 
Overall, the 
phase accumulation procedure in \texttt{BootMMPM} has
integer multiplication complexity $12nl_BNr\log N$.

In contrast, in TFHE, each CMux requires 2 multiplications between
RLWE ciphertext and RGSW ciphertext, where the polynomial ring is
${\cal R}_{Nr,Q}$. So the phase accumulation procedure in TFHE 
has integer multiplication complexity $12nl_BNr\log (Nr)$. 
In the setting of (\ref{set:par}), this complexity is 
$d_r$ times larger than that of Algorithm \texttt{BootMMPM}.

\vskip .2cm
Experiments:

We implement \texttt{BootMMPM} on Palisade platform\cite{palisade} and run it on two different
hardware platforms. We use the identity function $f(x)=x\in\mathbb{Z}_t$ 
for bootstrapping test. This function is not nega-cyclic, 
so two rounds of bootstrapping are required, with
the first extracting the MSB homomorphically. The first bootstrapping can be replaced by 
the homomorphic sign computing algorithm in \cite{liu2021large} to improve efficiency.
For the purpose of making comparison
with TFHE scheme, we adopt the same algorithm in both rounds of bootstrapping,
namely, either both are TFHE, or both are \texttt{BootMMPM}.

Platform 1. IBM-Compatible PC with Intel(R) Core(TM)
i7-10700 CPU @ 2.90 GHz and 16.0 GB RAM. 
Software: PALISADE v1.11.9 with compiler g++ 11.3.0.
Plaintext size: $\log t=5,6,\ldots,11$ respectively.
Following the parameter setting in \cite{liu2021large}, we choose
\be
\log n=9,\ \ \, \log Q=54, \ \ \, \log B=15,\ \ \, 
B_{\rm KS}=25.
\ee

\begin{table}[!t]
	\renewcommand{\arraystretch}{1.3}
	\caption{Experiments on Platform 1 (PC), with run-time and
		size of BootKeys (bootstrapping keys) and size of
		KSKeys (key-switching keys)}
	\label{table-exper}
	\centering
	\begin{tabular}{|c|c|c|c|c|c|c|}
		\hline
		& $\log t$ & $\log q$ & $\log N$ & $\log r$ & \makecell{average \\time (s) } & 
		\makecell{BootKeys \\
			+KSKeys} \\
		\hline
		\multirow{4}{*}{
			TFHE} & 5 & 12 & 11 & \multirow{4}{*}{} & 0.6969 &
		\makecell{256 MB\\+2.35 GB}
		\\
		\cline{2-4}\cline{6-7}
		& 6 & 13 & 12 & & 3.6953 & 
		\makecell{512 MB\\+4.70 GB}   
		\\
		\cline{2-4}\cline{6-7}
		& 7 & 14 & 13 & &10.092 & 
		\makecell{1 GB\\+9.39 GB}   
		\\
		
		\hline
		\multirow{7}{*}{\makecell{\texttt{Boot-}\\
				\texttt{MMPM}}} & 5 & 12 & \multirow{7}{*}{11} & 0 & 0.7375 &
		\multirow{7}{*}{\makecell{256 MB\\+2.35 GB}} \\
		\cline{2-3}\cline{5-6}
		& 6 & 13 &  &  1 & 1.3625 &  \\
		\cline{2-3}\cline{5-6}
		& 7 & 14 &  &  2 & 2.6376 &  \\
		\cline{2-3}\cline{5-6}
		& 8 & 15 &  & 3 & 5.2578 &  \\
		\cline{2-3}\cline{5-6}
		& 9 & 16 &  & 4 & 10.203 &  \\
		\cline{2-3}\cline{5-6}
		& 10 & 17 &  & 5 & 20.261 &  \\
		\cline{2-3}\cline{5-6}
		& 11 & 18 &  & 6 & 41.795 &  \\
		\hline
	\end{tabular}
\end{table}

In Table 1, the TFHE bootstrapping program
stops running when the plaintext size reaches 8 bits, with nothing coming out;
on the other hand, the \texttt{BootMMPM} program still runs 
when the plaintext size reaches 11 bits, and
finishes in 42 seconds with correct result.
When the plaintext has 7 bits, running TFHE costs over 10 seconds, while running
\texttt{BootMMPM} costs less than 3 seconds; the latter is faster by about 3 times.
For each plaintext size, 10 plaintexts of the specific size are randomly generated, and 
the algorithms run on the ciphertexts encrypting them separately to get the average 
run-time.

As to the key size, it can be reduced on the client's side using packing techniques
\cite{kim2021general,kim2023lfhe}, for both TFHE and \texttt{BootMMPM}, with the same scale. 
In the experiments, these improvements are not implemented for sharper comparison.

Platform 2. HP Workstation with Intel(R) Xeon(R) 
Gold 6258R CPU @ 2.70 GHz and 1.00 TB RAM. 
Software: PALISADE v1.11.8 with compiler g++ 11.3.0.
Plaintext size: $\log t=5,6,\ldots,15$ respectively.
Following \cite{albrecht2021homomorphic}, we increase $\log n$ to $10$.

Although the security level of \texttt{BootMMPM} under the above setting is lower than
that of TFHE when $r>1$, the increase of security level in TFHE is passively driven
by the need of larger exponent space, so that the whole look-up table can be encoded. 

\begin{table}[!t]
	\renewcommand{\arraystretch}{1.3}
	\caption{Experiments on Platform 2 (workstation)}
	\label{table-exper-2}
	\begin{center}
		\begin{tabular}{|c|c|c|c|c|c|c|}
			\hline
			& $\log t$ & $\log q$ & $\log N$ & $\log r$ & \makecell{average \\time (s) } & 
			\makecell{BootKeys \\
				+KSKeys} \\
			\hline
			\multirow{9}{*}{
				TFHE}& 5 & 12 & 11 &  & 1.7382 & \makecell{512 MB\\+4.69 GB} \\
			\cline{2-4}\cline{6-7}
			& 6 & 13 & 12 &  & 3.5343 & \makecell{1 GB\\+9.38 GB} \\
			\cline{2-4}\cline{6-7}
			& 7 & 14 & 13 & \multirow{5}{*}{} & 7.2890 &  \makecell{2 MB\\+18.8 GB}  \\
			\cline{2-4}\cline{6-7}
			& 8 & 15 & 14 &  & 14.894 & \makecell{4 GB\\+37.5 GB}  \\
			\cline{2-4}\cline{6-7}
			& 9 & 16 & 15 &  & 30.678 & \makecell{8 GB\\+75.1 GB} \\
			\cline{2-4}\cline{6-7}
			& 10 & 17 & 16 &  & 61.116 & \makecell{16 GB\\+150 GB} \\
			\cline{2-4}\cline{6-7}
			& 11 & 18 & 17 &  & 123.47 & \makecell{32 GB\\+300 GB}  \\
			\hline
			
			\multirow{9}{*}{\makecell{\texttt{Boot-}\\
					\texttt{MMPM}}
			} & 5 & 12 & \multirow{11}{*}{11} & 0 & 1.7687 & \multirow{9}{*}{\makecell{512 MB\\+4.69 GB}} \\
			\cline{2-3}\cline{5-6}
			& 6 & 13 & & 1 & 3.3859 &  \\
			\cline{2-3}\cline{5-6}
			& 7 & 14 & & 2 & 6.4064 &  \\
			\cline{2-3}\cline{5-6}
			& 8 & 15 & & 3 & 12.402 &  \\
			\cline{2-3}\cline{5-6}
			& 9 & 16 & & 4 & 24.379 &  \\
			\cline{2-3}\cline{5-6}
			& 10 & 17 &  & 5 & 49.792 &  \\
			\cline{2-3}\cline{5-6}
			& 11 & 18 & & 6 & 96.132 &  \\
			\cline{2-3}\cline{5-6}			
			& 12 & 19 & & 7 & 202.56 & \\
			\cline{2-3}\cline{5-6}
			& 13 & 20 & & 8 & 413.22 &  \\
			\cline{2-3}\cline{5-6}
			& 14 & 21 & & 9 & 834.60 &  \\
			\cline{2-3}\cline{5-6}
			& 15 & 22 &  & 10 & 1676.3 &  \\			
			\hline
		\end{tabular}
	\end{center}
\end{table}

In Table 2, both programs run correctly for plaintext size up to 11 bits, and
\texttt{BootMMPM} is slightly
faster. For example, when the plaintext size is
11 bits, TFHE costs more than 120 seconds, while \texttt{BootMMPM} costs less than 100 seconds.

On both platforms, \texttt{BootMMPM} remains constant bootstrapping key size 512 MB
and key-switching key size 4.69 GB,
while TFHE requires the bootstrapping key size to grow from 512 MB to 32 GB,
and the key-switching key size to grow from 4.69 GB to 300 GB.

\section{Conclusion}
For general functional bootstrapping of ciphertexts encrypting long plaintext,
this paper proposes to encode the look-up table of the function by a polynomial
vector, instead of only a polynomial in FHEW/TFHE. This motivates a thorough
investigation of the group of monic monomial permutation matrices and its cyclic
subgroups, based on which
a new algorithm for general functional bootstrapping is proposed. The algorithm 
features in small bootstrapping and key-switching key size, which benefits 
communication cost and memory cost in bootstrapping.

On the other hand, the new algorithm provides only slight improvement in time cost.
One may consider using a multivariate polynomial instead of a polynomial
vector to encode the look-up table of a general function. This idea may lead to significant
speed-up, just as in the case of parallel FHEW/TFHE bootstrapping 
\cite{liu2023batch1,liu2023batch2}.
New explorations are expected to make improvement on time cost in this direction.

\section*{Acknowledgments}
This research was supported partially by China National Key Research and Development Project 
Grant No. 2020YFA0712300.


%

\bibliographystyle{IEEEtran}

\bibliography{bmyref.bib}

\begin{thebibliography}{10}
\providecommand{\url}[1]{#1}
\csname url@samestyle\endcsname
\providecommand{\newblock}{\relax}
\providecommand{\bibinfo}[2]{#2}
\providecommand{\BIBentrySTDinterwordspacing}{\spaceskip=0pt\relax}
\providecommand{\BIBentryALTinterwordstretchfactor}{4}
\providecommand{\BIBentryALTinterwordspacing}{\spaceskip=\fontdimen2\font plus
\BIBentryALTinterwordstretchfactor\fontdimen3\font minus
  \fontdimen4\font\relax}
\providecommand{\BIBforeignlanguage}[2]{{%
\expandafter\ifx\csname l@#1\endcsname\relax
\typeout{** WARNING: IEEEtran.bst: No hyphenation pattern has been}%
\typeout{** loaded for the language `#1'. Using the pattern for}%
\typeout{** the default language instead.}%
\else
\language=\csname l@#1\endcsname
\fi
#2}}
\providecommand{\BIBdecl}{\relax}
\BIBdecl

\bibitem{tan2020efficient}
B.~H.~M. Tan, H.~T. Lee, H.~Wang, S.~Ren, and K.~M.~M. Aung, ``Efficient
  private comparison queries over encrypted databases using fully homomorphic
  encryption with finite fields,'' \emph{IEEE Transactions on Dependable and
  Secure Computing}, vol.~18, no.~6, pp. 2861--2874, 2020.

\bibitem{yi2012single}
X.~Yi, M.~G. Kaosar, R.~Paulet, and E.~Bertino, ``Single-database private
  information retrieval from fully homomorphic encryption,'' \emph{IEEE
  Transactions on Knowledge and Data Engineering}, vol.~25, no.~5, pp.
  1125--1134, 2012.

\bibitem{angel2018pir}
S.~Angel, H.~Chen, K.~Laine, and S.~Setty, ``Pir with compressed queries and
  amortized query processing,'' in \emph{2018 IEEE symposium on security and
  privacy (SP)}.\hskip 1em plus 0.5em minus 0.4em\relax IEEE, 2018, pp.
  962--979.

\bibitem{cong2022sortinghat}
K.~Cong, D.~Das, J.~Park, and H.~V. Pereira, ``Sortinghat: Efficient private
  decision tree evaluation via homomorphic encryption and transciphering,'' in
  \emph{Proceedings of the 2022 ACM SIGSAC Conference on Computer and
  Communications Security}, 2022, pp. 563--577.

\bibitem{gentry2009fully}
C.~Gentry, ``Fully homomorphic encryption using ideal lattices,'' in
  \emph{Proceedings of the forty-first annual ACM symposium on Theory of
  computing}, 2009, pp. 169--178.

\bibitem{alperin2014faster}
J.~Alperin-Sheriff and C.~Peikert, ``Faster bootstrapping with polynomial
  error,'' in \emph{Annual Cryptology Conference}.\hskip 1em plus 0.5em minus
  0.4em\relax Springer, 2014, pp. 297--314.

\bibitem{hiromasa2016packing}
R.~Hiromasa, M.~Abe, and T.~Okamoto, ``Packing messages and optimizing
  bootstrapping in gsw-fhe,'' \emph{IEICE TRANSACTIONS on Fundamentals of
  Electronics, Communications and Computer Sciences}, vol.~99, no.~1, pp.
  73--82, 2016.

\bibitem{ducas2015fhew}
L.~Ducas and D.~Micciancio, ``Fhew: bootstrapping homomorphic encryption in
  less than a second,'' in \emph{Annual international conference on the theory
  and applications of cryptographic techniques}.\hskip 1em plus 0.5em minus
  0.4em\relax Springer, 2015, pp. 617--640.

\bibitem{chillotti2016faster}
I.~Chillotti, N.~Gama, M.~Georgieva, and M.~Izabachene, ``Faster fully
  homomorphic encryption: Bootstrapping in less than 0.1 seconds,'' in
  \emph{international conference on the theory and application of cryptology
  and information security}.\hskip 1em plus 0.5em minus 0.4em\relax Springer,
  2016, pp. 3--33.

\bibitem{liu2021large}
Z.~Liu, D.~Micciancio, and Y.~Polyakov, ``Large-precision homomorphic sign
  evaluation using fhew/tfhe bootstrapping,'' \emph{Cryptology ePrint Archive},
  2021.

\bibitem{chillotti2021improved}
I.~Chillotti, D.~Ligier, J.-B. Orfila, and S.~Tap, ``Improved programmable
  bootstrapping with larger precision and efficient arithmetic circuits for
  tfhe,'' in \emph{Advances in Cryptology--ASIACRYPT 2021: 27th International
  Conference on the Theory and Application of Cryptology and Information
  Security, Singapore, December 6--10, 2021, Proceedings, Part III 27}.\hskip
  1em plus 0.5em minus 0.4em\relax Springer, 2021, pp. 670--699.

\bibitem{chillotti2020tfhe}
I.~Chillotti, N.~Gama, M.~Georgieva, and M.~Izabach{\`e}ne, ``Tfhe: fast fully
  homomorphic encryption over the torus,'' \emph{Journal of Cryptology},
  vol.~33, no.~1, pp. 34--91, 2020.

\bibitem{yang2021tota}
Z.~Yang, X.~Xie, H.~Shen, S.~Chen, and J.~Zhou, ``Tota: Fully homomorphic
  encryption with smaller parameters and stronger security,'' \emph{Cryptology
  ePrint Archive}, 2021.

\bibitem{liu2023batch1}
F.-H. Liu and H.~Wang, ``Batch bootstrapping i: a new framework for simd
  bootstrapping in polynomial modulus,'' in \emph{Annual International
  Conference on the Theory and Applications of Cryptographic Techniques}.\hskip
  1em plus 0.5em minus 0.4em\relax Springer, 2023, pp. 321--352.

\bibitem{liu2023batch2}
{Liu, Feng-Hao and Wang, Han}, ``Batch bootstrapping ii: bootstrapping in
  polynomial modulus only requires o\~{}(1) fhe multiplications in
  amortization,'' in \emph{Annual International Conference on the Theory and
  Applications of Cryptographic Techniques}.\hskip 1em plus 0.5em minus
  0.4em\relax Springer, 2023, pp. 353--384.

\bibitem{lee2023efficient}
Y.~Lee, D.~Micciancio, A.~Kim, R.~Choi, M.~Deryabin, J.~Eom, and D.~Yoo,
  ``Efficient fhew bootstrapping with small evaluation keys, and applications
  to threshold homomorphic encryption,'' in \emph{Annual International
  Conference on the Theory and Applications of Cryptographic Techniques}.\hskip
  1em plus 0.5em minus 0.4em\relax Springer, 2023, pp. 227--256.

\bibitem{kim2021general}
A.~Kim, M.~Deryabin, J.~Eom, R.~Choi, Y.~Lee, W.~Ghang, and D.~Yoo, ``General
  bootstrapping approach for rlwe-based homomorphic encryption,''
  \emph{Cryptology ePrint Archive}, 2021.

\bibitem{kim2023lfhe}
A.~Kim, Y.~Lee, M.~Deryabin, J.~Eom, and R.~Choi, ``Lfhe: Fully homomorphic
  encryption with bootstrapping key size less than a megabyte,''
  \emph{Cryptology ePrint Archive}, 2023.

\bibitem{palisade}
``Palisade lattice cryptography library,'' \url{https://palisade-crypto.org/},
  2022.

\bibitem{albrecht2021homomorphic}
M.~Albrecht, M.~Chase, H.~Chen, J.~Ding, S.~Goldwasser, S.~Gorbunov, S.~Halevi,
  J.~Hoffstein, K.~Laine, K.~Lauter \emph{et~al.}, ``Homomorphic encryption
  standard,'' \emph{Protecting privacy through homomorphic encryption}, pp.
  31--62, 2021.

\end{thebibliography}

\begin{IEEEbiographynophoto}{Dengfa Liu}
	received Bachelor of Science degree from Fujian Normal
	University in 2019. He is currently pursuing a PhD at AMSS,
	Chinese Academy of Sciences.
	His research interest is Algorithm Design in Privacy Computation.
\end{IEEEbiographynophoto}

\begin{IEEEbiographynophoto}{Hongbo Li}
	is Kwan Chao-Chi Chair Professor and Director
	of the Institute of Systems Science, AMSS, Chinese Academy of Sciences.
	He received BS, MS, PhD from the Department of Mathematics at 
	Beijing University in 1988, 1991, 1994, respectively. His research interests
	include Automated Reasoning, Symbolic Computation,
	Geometric Algebra, Quantum Computing, 
	Privacy Computation, etc. He is Director of the Computer Mathematics Committee
	of Chinese Mathematics Society.
	
\end{IEEEbiographynophoto}

\vfill

\end{document}